\documentclass[DIV=20,10pt,a4paper,twocolumn,headings=normal]{scrartcl}

\usepackage[english]{babel}
\usepackage{graphicx}					
\usepackage{amsmath}
\usepackage{amssymb}
\usepackage{bm}	
\usepackage{abstract} 
\usepackage{flushend} 
\usepackage{cuted} 
\usepackage{mathenv} 
\usepackage[title]{appendix} 
\usepackage{cite}

\usepackage[unicode, pdfstartview= FitH]{hyperref} 

\begin{document}

\renewcommand{\abstractname}{\vspace{-50pt}} 
\renewcommand{\refname}{\vspace{-23pt}}  

\title{Statistical mechanics of fluids at an impermeable wall}

\author{V. M. Zaskulnikov  \thanks{zaskulnikov@gmail.com}
\\Institute of Chemical Kinetics and Combustion, 
\\Institutskaya, 3,  Novosibirsk, 630090, Russian Federation}

\date{\today}

\twocolumn[

\maketitle

\begin{onecolabstract}

The problem of surface effects at a fluid/force field boundary is investigated. A classical simple fluid with a locally introduced field simulating a solid is considered. For the case of a hard-core field, rigid, exponential, realistic, and macroscopically smooth boundaries are examined.

Two approaches to this problem are analyzed, which have been developed independently from each other over a long period of time. With some degree of arbitrariness, they can be referred to as ``adsorption'' vs ``surface tension'' or ``cluster expansion'' vs ``pressure tensor''.

The ``adsorption'' approach is used to obtain a series in powers of the activity for the specific surface omega potential (grand potential) - $\gamma$. This expression is similar to the Mayer expansion, but, unlike in the case of pressure, the integrals of the Ursell functions contain factors which depend on the particle/wall interaction potential. These factors are written in the general form for an arbitrary term of the series, and have a fairly simple structure. In the case of a hard wall, the coefficients of the series reduce to the first moments of the Ursell functions taken over certain regions. Equivalence of the ``adsorption'' and the ``local pressure'' approaches is established.

The ``surface tension'' approach is used to expand the Kirkwood-Buff formula for the surface omega potential of the field/fluid system to the arbitrary localization of the dividing surface. The surface terms identification depends on the particular form of the fluid particle/solid interaction potential, and on the formulation of the problem. As a function of an activity, the surface terms can begin both with linear term and quadratic one, depending on the situation, and this corresponds to the presence or absence of adsorption in the ordinary sense. ``The surface tension coefficient'' breaks up into the term proportional to the Henry constant, depending on the dividing surface position, and universal nonlinear surface coefficient.
  
Using methods of statistical mechanics, it is shown that the derivative of the tangential component of the pressure tensor with respect to the chemical potential coincides with the near-surface number density on average over the transition region, that has two consequences.

Firstly, it proves complete identity between ``tension'' and ``adsorption'' approaches in the domain of their existence. Thus, the results of one line of investigation can be used to develop the other. In particular, the ``adsorption'' method obtains the expression for the surface terms in quadratures, and the ``tension'' technique - the analog of the Mayer expansion. 

Secondly, it gives the near-surface virial expansion, which determines the exact equation of state of near boundary   ``two-dimensional'' fluid. The tangential component of the pressure tensor averaged over the transition region plays the role of pressure, and the average number density - the role of number density.

\vspace{20pt}

\end{onecolabstract}

]

\saythanks

\section{\label{sec:01}Introduction}

The surface terms at the solid/fluid interface for the case of ``gas adsorption on a solid surface'' have been studied in terms of the statistical mechanics for a fairly long time \cite{Ono1950, hill1959, OnoKondo1960, hillstatmeh1987}. Most papers consider a solid as a constant field, as the first step of a more complex problem. Here we will not deal with the problems related to the motion of solid particles. 

More correct calculation \cite{Bellemans1962}, as shown later, corresponds to a hard solid. Attempts were made to combine these results \cite{Bakri1966}.

In subsequent studies, the initial consideration \cite{Bellemans1962} was extended to the case of an arbitrary particle/surface interaction potential \cite{SteckiSokolowski1978, SokolowskiStecki1980, SokolowskiStecki1981}. Inaccuracies of early studies \cite{OnoKondo1960, hillstatmeh1987} related to the integration domains of the series terms were also corrected \cite{SokolowskiStecki1980, SokolowskiStecki1981}.

A significant drawback of this approach \cite{Bellemans1962, SokolowskiStecki1980} is that the computation algorithm is extremely complex or, in other words, there is no universal expression for the general term of the series.

The second problem is related to the Henry constant, and here we observe three cases.

In the most exotic one, the interface locates at the first layer of solid atoms \cite{OnoKondo1960, hillstatmeh1987, SokolowskiStecki1980, SokolowskiStecki1981}.

In some studies \cite{Bellemans1962} this constant is strictly zero, while in many other papers, it is proportional to the excluded volume \cite{hill1959, MeelCharbonneau2009, LairdDavidchak2010}.

The equations used in some papers \cite{hill1959}, \cite{RenHsue2003} give negative values for the number of surface particles under certain conditions.

These problems are due to the fact that the technique of arbitrary localization of the dividing surface has not been clarified yet \cite{SokolowskiStecki1980, SokolowskiStecki1981}.

This question is closely related to the issue of interpretation of the surface number density, so, it is also not fully resolved.

Another reference point in the interface problem is the Kirkwood-Buff formula \cite{KirkwoodBuff1949} for the surface tension of the liquid/gas system
\begin{equation}
\gamma  =  \int \left [ P- P'(x)  \right ] dx. 
\label{eq:001}
\end{equation}

Here $P$ is the pressure in the homogeneous phases, $P'$ is interpreted as the tangential component of the pressure tensor \cite{IrvingKirkwood1950, Harasima1958}, and $x$ is the coordinate normal to the interface.

In addition to the force-based consideration \cite{KirkwoodBuff1949}, a similar formula was derived earlier using a quasi-thermodynamic analysis \cite{Tolman1948}. In this case $P'$ has the meaning of the equilibrium pressure corresponding to the number density in the transition region.

Equations such as (\ref{eq:001}) are also applied to the solid (field)/fluid interface \cite{HendersonSwol1984, Navascues1979, LairdDavidchak2010}. (In some cases, the normal component of the pressure tensor - $P_n$ was extracted from $P$ \cite{NavascuesBerry1977, HendersonSwol1984, LairdDavidchak2010}. In terms of further consideration, this manipulation seems meaningless.)

An application of this equation to the field/fluid interface, in our opinion, requires further validation.

Particularly the question of integration limits in equation (\ref{eq:001}) is not quite clear. In the case of liquid/gas systems the integration is performed from $-\infty$ to $+\infty$, however, it is impossible for solid (field)/fluid systems due to $P'$ zeroing inside the solid phase. For some reasons, possible variants of generalization \cite{StillingerBuff1962} have not received logical continuation. Thus, here we also have the above-mentioned problem of the dividing surface localization. 

The ``mechanical'' consideration (section \ref{subsec:03j}) has one significant drawback: it deals with the tangential pressure which is distorted by the transverse wall (section \ref{subsec:04c}). This issue is not adequately covered in the literature \cite{HendersonSwol1984}.

Finally, from the viewpoint of analytical space expansion, and the improvement of the problem ``ecology'', it is of interest to establish whether the ``adsorption'' direction \cite{Ono1950, hill1959, OnoKondo1960, hillstatmeh1987, Bellemans1962, SokolowskiStecki1980, SokolowskiStecki1981} is equivalent to ``surface tension'' direction \cite{KirkwoodBuff1949, Harasima1958, Navascues1979, HendersonSwol1984, SchofieldHenderson1982} concerning force field/fluid interface.

Some results, related to this contribution were obtained in our previous work \cite{zaskulnikov201004a}. In particular, this regards to the technique of partial localization factors, which generalize the notions of Boltzmann and Urcell factors.

\section{\label{sec:02}Primary definitions}

\subsection{\label{subsec:02a}Canonical ensemble}

The probability density of finding a given spatial configuration of a particular set of particles \cite[p.181]{hillstatmeh1987} is given by the expression
\begin{equation}
P^{(k)}_{1...k} = \frac{1}{Z_N}\int \limits_V \exp(-\beta U^N_{1...N})d\bm{r}_{k+1}\dots d\bm{r}_N. 
\label{eq:002}
\end{equation}

Here $N$ is the number of particles in the system, $\beta = 1/k_BT$, $k_B$ is the Boltzmann constant, $T$ is the temperature, $U^N_{1...N}$ is the energy of interaction between the particles, and $V$ is the volume of the system. The integration is performed over the coordinates of the particles of the ensemble $\bm{r}_{k+1},\dots,\bm{r}_N$. $Z_N$ is the configuration integral
\begin{equation}
Z_N = \int \limits_V \exp(-\beta U^N_{1...N}) d\bm{r}_1...d\bm{r}_N. 
\label{eq:003}
\end{equation}

Going over to the distribution functions for an arbitrary set of particles, we obtain
\begin{equation}
\varrho^{(k)}_{C,1...k} = \frac{N!}{(N-k)!}P^{(k)}_{1...k}, 
\label{eq:004}
\end{equation}
where $\varrho^{(k)}_{C, 1 ... k}$ is the probability density of finding a given configuration of $k$ arbitrary particles for the canonical ensemble.

\subsection{\label{subsec:02b}Grand canonical ensemble (GCE)}

Let us average equation (\ref{eq:004}) over fluctuations in the number of particles by applying to both of its sides the operation $\sum_{N=0}^\infty P_N^V$, where
\begin{equation}
P_N^V = \frac{z^N Z_N}{N!\Xi_V} 
\label{eq:005}
\end{equation}
is the probability for the GCE to have a definite number of particles $N$ inside the volume $V$. Here $z$ is the activity:
\begin{equation}
z = \frac{e^{\mu/k_BT}}{\Lambda^3}, 
\label{eq:006}
\end{equation}
where $\mu$ is the chemical potential, $\Lambda = h/\sqrt[]{2 \pi mk_BT}$, $h$ is Planck's constant, $m$ is the particle mass, and $\Xi_V$ is the large partition function of a system of volume $V$: 
\begin{equation}
\Xi_V = 1 + \sum_{N=1}^\infty \frac{z^N Z_N}{N!}. 
\label{eq:007}
\end{equation}

We have
\begin{equation}
\varrho^{(k)}_{G,1...k} = \sum_{N=k}^\infty \varrho^{(k)}_{C,1...k} P_N^V,
\label{eq:008}
\end{equation}
or 
\begin{eqnarray}
\varrho^{(k)}_{G,1...k}  =&& \frac{z^k}{\Xi_V}  \Big \{ \exp(-\beta U^{k}_{1...k})  +  \sum_{N=1}^\infty \frac{z^N}{N!} \label{eq:009} \\
&&    \times  \int \limits_V   \exp(-\beta U^{N+k}_{1...N+k})d\bm{r}_{k+1}...d\bm{r}_{k+N} \Big \}  \nonumber.
\end{eqnarray}

$\varrho^{(k)}_{G,1...k}$ specify the probability density of finding a certain configuration of $k$ arbitrary particles in the GCE. For an ideal gas, $\varrho^{(k)}_{G,1...k} = \varrho^{k}$, where $\varrho = \overline{N}/V$ is the number density.

\subsection{\label{subsec:02c}Ursell factors and partial localization factors}

The Ursell factors ${\cal U}^{(k)}_{1...k}$, which are also called cluster functions, appear in the well-known expansion of the pressure in powers of the activity \cite[p.129]{hillstatmeh1987}, \cite[p.232]{landaulifshitz1985}
\begin{equation}
P(z,T) =zk_BT  + k_BT \sum_{k=2}^\infty \frac{z^k}{k!} \int {\cal U}^{(k)}_{1...k} d\bm{r}_2...d\bm{r}_k.
\label{eq:010} 
\end{equation}

They are also contained in the corresponding expansion of the number density
\begin{equation}
\varrho(z) = z + z\sum_{n=1}^\infty \frac{z^n}{n!} \int {\cal U}^{(n+1)}_{1...n+1} d\bm{r}_{2}...d\bm{r}_{n+1},
\label{eq:011}
\end{equation}
which is obvious from the relation \footnote[1]{To simplify the formulae, we do not indicate temperature constancy; it will be assumed in all derivatives. In addition, differentiation parameters will not be indicated in cases where this is obvious, for example, when they are set by the opposite side of the equation.}
\begin{equation}
\varrho = \frac{\partial P}{\partial \mu}.
\label{eq:012}
\end{equation}

The Ursell factors decay rapidly as any group of particles, including a single particle, moves away.

Because (\ref{eq:010}) is, in fact, a Taylor series expansion and in view of the local nature of the Ursell factors, we can write the equation
\begin{equation}
{\cal U}^{(k)}_{1...k}  = \beta \frac{\partial^{k}P}{\partial z^{k} } \prod_{n = 2}^k \delta(\bm{r}_n - \bm{r}_1),
\label{eq:013}
\end{equation}
which is valid on macroscopic scales, and where the derivatives are taken at the point $z = 0$. Here $\delta(\bm{r})$ is the Dirac delta function. Naturally, in this case, we assume that the index $k$ does not reach macroscopic values.

Some other properties of the Ursell factors are described in Appendix \ref{subsec:appenda1}.

The partial localization factors ${\cal B}^{(m,k)}_{1...m+k}$ are the hybrids of the Boltzmann and Ursell factors. They are contained in the expansion of the distribution functions in powers of the activity
\begin{eqnarray}
\varrho^{(m)}_{G,1...m} (\psi^V)&& = z^m \Big \{ {\cal B}^{(m,0)}_{1...m}  +  \sum_{k=1}^\infty \frac{z^k}{k!} \label{eq:014} \\
 \times &&   \int   \Big [ \prod_{i = m+1}^{m+k} \psi^V_i \Big ]   {\cal B}^{(m,k)}_{1...m+k} d\bm{r}_{m+1}...d\bm{r}_{m+k} \Big \}  \nonumber,
\end{eqnarray}
where
\begin{equation}
\psi^{V}_i = \psi^{V}(\bm{r}_i) = 
	 \left\{ 
			\begin{array}{ll} 
         1 & (\bm{r}_i \in V)\\   
         0 & (\bm{r}_i \notin V)
     	\end{array}  
		\right.
			\label{eq:015}
\end{equation}
are  characteristic functions. 

The properties of ${\cal B}^{(m,k)}_{1...m+k}$ are described in Appendix \ref{subsec:appenda2}.

\subsection{\label{subsec:02d}Presence of an external field}

The configuration integral of the inhomogeneous closed system is given by the expression
\begin{equation}
Z^U_N = \int\limits_{V}^{} \exp(-\beta\sum_{i=1}^N v_i-\beta U^N_{1...N}) d\bm{r}_1...d\bm{r}_N, \\
\label{eq:016}
\end{equation}
where $v_i$ is the energy of the interaction of the $i$-th particle with the field.

For the GCE, we introduce the quantity
\begin{equation}
\Xi^U_V = 1+ \sum_{N=1}^\infty \frac{z^N Z^U_N}{N!}, 
\label{eq:017}
\end{equation}
which is obviously the large partition function of the system in the presence of an external field.

Using the properties of the fractional generating function (\ref{eq:009}) (see Appendices \ref{subsec:appenda2}, \ref{subsec:appendb1}), we obtain the analog of (\ref{eq:011})
\begin{equation}
\varrho(\bm{r_1},z) = z \theta_1 + z\theta_1\sum_{n=1}^\infty \frac{z^n}{n!} \int \Big [ \prod_{i = 2}^{n+1} \theta_i \Big ]{\cal U}^{(n+1)}_{1...n+1} d\bm{r}_{2}...d\bm{r}_{n+1},
\label{eq:018}
\end{equation}
where $\varrho(\bm{r_1},z)$ is the number density in the presence of an external field and
\begin{equation}
\theta_i =  \exp(-\beta v_i) = \exp[-\beta v(\bm{r_i})],
\label{eq:019}  
\end{equation}
and the analog of (\ref{eq:014})
\begin{eqnarray}
\varrho^{(m)}_{1...m} (\theta)&& = z^m \Big  [ \prod_{i = 1}^{m} \theta_i \Big  ] \Big  \{ {\cal B}^{(m,0)}_{1...m}  +  \sum_{k=1}^\infty \frac{z^k}{k!}  \label{eq:020} \\
&&   \times  \int  \Big  [ \prod_{i = m+1}^{m+k} \theta_i \Big  ]   {\cal B}^{(m,k)}_{1...m+k} d\bm{r}_{m+1}...d\bm{r}_{m+k} \Big \}.  \nonumber
\end{eqnarray}

\section{\label{sec:03}System in a force field}

We consider a statistical system in a field whose range is much smaller than the size of the system and which is far from the boundaries of the system. That is, the case in point is rather ``the field in the system''.

We also assume that the potential of the fields considered increases/decays rapidly enough to ensure the convergence of the corresponding integrals for the motion inward and outward the solid, and the interparticle potential decays rapidly enough to provide the convergence of zero and first moments of the Ursell factors (i.e., we do not consider the Coulomb potentials in this paper).

In some cases, we assume that the range of the solid field has a finite radius, so we speak about a transition region between the solid and the fluid. This approach is of no fundamental nature and will be used to simplify the consideration.

Finally, unless otherwise stated, we assume that the region of the transition from the solid to the fluid is much smaller than the linear sizes of the region of the field, and that the field profile along its gradient is the same everywhere.

\subsection{\label{subsec:03a}Basic equations}

Taking the logarithm of the ratio of the partition functions (\ref{eq:017}) and (\ref{eq:007}) gives
\begin{eqnarray}
 \ln{\Xi^U_V} -&&  \ln{\Xi_V} = \sum_{k=1}^\infty \frac{z^k}{k!} \label{eq:021} \\
&& \int\limits_{V}^{}\Big [\exp(-\beta\sum_{i=1}^k v_i)-1 \Big ] {\cal U}^{(k)}_{1...k} d\bm{r}_1...d\bm{r}_k
\nonumber,
\end{eqnarray}
where the logarithmic form of the generating function for the Ursell factors is used.

Under the above assumptions, we can obviously omit the symbol $V$ in the integrals over the coordinates of particles in (\ref{eq:021}) and consider the integration to be performed over the infinite space.

We rearrange (\ref{eq:021}) as follows:
\begin{eqnarray}
 \ln{\Xi^U_V}&&-\ln{\Xi_V} = - \sum_{k=1}^\infty \frac{z^k}{k!} \int \varphi_1 {\cal U}^{(k)}_{1...k} d\bm{r}_1...d\bm{r}_k \label{eq:022}  \\
&& + \sum_{k=2}^\infty \frac{z^k}{k!} \int \theta_1 \Big [ \prod_{l = 2}^k (1-\varphi_l) 
 - 1  \Big ]{\cal U}^{(k)}_{1...k} d\bm{r}_1...d\bm{r}_k
\nonumber,
\end{eqnarray}
where we use the notation
\def\bydefn{\stackrel{def}{=}}\def\convf{\hbox{\space \raise-2mm\hbox{$\textstyle      \bigotimes \atop \scriptstyle \omega$} \space}}
\begin{equation}
\varphi_i \bydefn \varphi(\bm{r}_i) = 1 - \exp(-\beta v_i) = 1 - \theta_i. \\
\label{eq:023}
\end{equation}

As can be seen from (\ref{eq:010}), the first sum in (\ref{eq:022}) is proportional to the Mayer series for the pressure of a homogeneous system.

The second term on the right side of (\ref{eq:022}) is proportional to the area of the surface which separates the field region from the rest of the system. Indeed, its structure is such that at least two particles are always on the opposite sides of the ``field boundary''. Expanding the product in any term of it, we have
\begin{equation}
\int \theta_1 \Big [ \prod_{i = 2}^{j} \varphi_i \Big ] {\cal U}^{(t)}_{1...t} d\bm{r}_1...d\bm{r}_t,
\label{eq:024}  
\end{equation}
where $2\leq j\leq t$. With the first particle fixed, the integrals of the type of (\ref{eq:024}) are determined by a local region near this particle because of the locality of the Ursell factors. Consequently, $\int ... d\bm{r}_2...d\bm {r}_t$ do not  depend on the displacement of the first particle along the boundary of the system. 

As the first particle moves from the boundary into the field region, the integrand of (\ref{eq:024}) decays rapidly due to the Boltzmann factor $\theta_1$.

As the first particle moves from the boundary into the fluid region, it decays rapidly because of the fixing factors $\varphi_i$ and the locality of the Ursell factors. We can say that the Ursell factor ``glues'' functions $\theta_1$ and $\varphi_i$, i.e., the fluid and field regions.

Integrating over the surface, factoring the area out of the sum, and after simple manipulations, we obtain 
\begin{equation}
\Omega^U = -PV +P\int \varphi d\bm{r} + \nu A,
\label{eq:025}
\end{equation}
where $\Omega^U$ is the omega potential (grand potential) of the system with the field embedded in it, $A$ is the area which bounds the field region, and we introduce the notation
\begin{equation}
\nu (z,T) =  k_BT \sum_{k=2}^\infty \frac{z^k}{k!}\int   \theta_1  \Big [1 - \prod_{l = 2}^k\theta_l \Big ]{\cal U}^{(k)}_{1...k} dx_1 d\bm{r}_2...d\bm{r}_k,
\label{eq:026}
\end{equation} 
where $x_1$ is the coordinate directed along the field gradient and it is chosen so as to satisfy the condition $dx_1> 0$.

Recall that the integration is performed over the infinite space.

The first terms of series, which is analogous to (\ref{eq:026}), were calculated in the set of papers \cite{SokolowskiStecki1980, SokolowskiStecki1981} based on a topological approach \cite{Bellemans1962} extended beyond the potential of a hard wall.

Since the Ursell factors depend only on the relative configuration of particles, in (\ref{eq:026}) we can perform integration over $dx_1$. Making the change of variables $\bm{r}_1^{'} = \bm{r}_1, \bm{r}_i^{'} = \bm{r}_i -\bm {r}_1, i = 2, \dots k$ and using the invariance of ${\cal U}^{(k)}_{1...k}$ under translations, we get
\begin{equation}
\nu (z,T) =  k_BT \sum_{k=2}^\infty \frac{z^k}{k!}\int {f}^{(k-1)}_{2...k}  {\cal U}^{(k)}_{0,2...k} d\bm{r}_2...d\bm{r}_k,
\label{eq:027}
\end{equation}
where
\begin{equation}
{f}^{(k-1)}_{2...k} = \int \limits_{- \infty}^{+ \infty} \theta(x_1)  \Big [1 - \prod_{i = 2}^k\theta(x_1 + x_i) \Big ] d x_1
\label{eq:028}
\end{equation}
is the function of the variables $x_2,... x_k$, which is symmetric under permutations of particles, and has the dimension of length. Obviously, in contrast to ${\cal U}^{(k)}_{1...k}$, it is not invariant with respect to translations since it depends on the location of the external field. 

Note that in (\ref{eq:027}) the variables are separated: the external potential is contained only in ${f}^{(k-1)}$ and the interparlicle one - only in ${\cal U}^{(k)}$. 

Expressions (\ref{eq:025}) and (\ref{eq:027}) are basic for the further consideration. The first term on the right side of (\ref{eq:025}) has purely bulk properties, and the third one has purely surface properties. The second term is of a mixed nature and has both bulk and surface properties, as is easily seen from its structure.

It is obvious from the foregoing that the attempts to represent the $\Omega$-potential as a sum of uniquely defined bulk and surface terms are, generally speaking, incorrect. There should be a free parameter - the position of the dividing surface.

The function $\nu$ plays an important role in the problem considered. Other representations of this quantity will be given below. Formula (\ref{eq:027}) is, in fact, an analog of the Mayer expansion (\ref{eq:010}), which, generally speaking, cannot be further simplified. Note that, as is easily seen, the expansion of the function $\nu (z,T)$ in a series of $z$ begins with a quadratic term. Thus, it is logical to call this value a nonlinear surface coefficient.

\subsection{\label{subsec:03b}Variants of accounting for surface effects}

In the derivation of (\ref{eq:026}) - (\ref{eq:028}), we assumed that the potential of the external field is independent of the displacement along the boundary. Suppose, moreover, that it has a hard core, and, for simplicity, in some cases we assume that the transition region is strongly restricted to a certain length.

We rearrange (\ref{eq:025}) as
\begin{equation}
\Omega = -P(V-V') - PA\int\limits_{-\infty}^{x'}\theta dx+ PA\int\limits_{x'}^{\infty}\varphi dx + \nu A,
\label{eq:029}
\end{equation}
where $x'$ is some arbitrary point within or near the transition layer that determines the volume $V'$, and $x$ is the coordinate directed along the field gradient. (With the same result, the integration region in equation (\ref{eq:029}) can be restricted to the transition layer because outside it the integrands vanish.)

We see that there is some degree of arbitrariness in the differentiation between bulk and surface terms. Actually, this arbitrariness exists only in the transition layer or near it. Indeed, it is clear that otherwise we have the banal compensation of two identical terms of opposite signs.

We denote the volume of the hard core by $V_1$, and the volume of the transition region by $V_t$. As noted above, $V_t\ll V_1$. 

From (\ref{eq:029}) we obtain at least three variants of accounting for surface effects for different values of $x'$:
\begin{equation}
\Omega = -P(V-V_1-V_t) - PA\int\limits_{l_t}^{} \theta dx + \nu A,
\label{eq:030}
\end{equation}
\begin{equation}
\Omega = -P(V-V_1) + PA\int\limits_{l_t}^{} \varphi dx + \nu A,
\label{eq:031}
\end{equation}
\begin{equation}
\Omega = -P(V-V_0) + \nu A,
\label{eq:032}
\end{equation}
where $l_t$ is the length of the transition region and $V_0$ is the volume of the solid bounded by the surface defined by the condition
\begin{equation}
\int\limits_{-\infty}^{x_0} \theta dx = \int\limits_{x_0}^{\infty} \varphi dx. 
\label{eq:033}
\end{equation}

On the left side of this equality, the integration is performed outward from the field region, and on the right side, inward to the fluid, and it determines the position of the surface in question $x_0$.

Variant (\ref{eq:030}) corresponds to $x'$ lying on the boundary of a homogeneous fluid, i.e., it assigns surface effects to the solid, while variant (\ref{eq:031}) corresponds to $x'$ lying on the boundary of a homogeneous solid and assigns surface effects to the fluid. Variant (\ref{eq:032}) sets the linear surface terms equal to zero, using the property (\ref{eq:033}). In this case, $x'$ can be both within and outside the transition region, depending on the form of the external potential.

In computational studies, one often uses the fourth variant in which the integration is extended to the inaccessible volume near the surface to keep the volume of the entire system. This example will be discussed below. Of course, the other variants based on formula (\ref{eq:029}) can also be considered.

The results obtained above will be discussed below, and here we present the traditional expression for (\ref{eq:029})
\begin{equation}
\Omega = -P(V-V')  + \gamma(x') A,
\label{eq:034}
\end{equation}
where
\begin{equation}
\gamma(x') = - P\int\limits_{-\infty}^{x'}\theta dx  + P\int\limits_{x'}^{\infty}\varphi dx + \nu 
\label{eq:035}
\end{equation}
is the general form of the specific surface $\Omega$- potential. In this expression, the term linear in pressure is separated from the nonlinear one - $\nu$.

\subsection{\label{subsec:03c}Surface number density}

Differentiating (\ref{eq:029}) with respect to the chemical potential, in view of  
\begin{equation}
N = - \left ( \frac{\partial \Omega}{\partial \mu} \right )_{V,A},   
\label{eq:036} 
\end{equation}
and using (\ref{eq:012}), we obtain
\begin{equation}
N = \varrho(V-V')+ \varrho A\int\limits_{-\infty}^{x'}\theta dx - \varrho A\int\limits_{x'}^{\infty}\varphi dx - A\beta z \frac{\partial\nu}{\partial z },
\label{eq:037}
\end{equation}
or
\begin{equation}
N = N_b  + N_s,
\label{eq:038}  
\end{equation}
where $N_b$ and $N_s$ are the number of bulk and surface particles, respectively. For them, we have 
\begin{equation}
N_b =  \varrho(V-V'),
\label{eq:039}  
\end{equation}
\begin{equation}
N_s =  \varrho A\int\limits_{-\infty}^{x'}\theta dx - \varrho A\int\limits_{x'}^{\infty}\varphi dx - A\beta z \frac{\partial\nu}{\partial z }.
\label{eq:040}  
\end{equation}

Going to the surface number density
\begin{equation}
\varrho_s =   \frac{N_s}{A},
\label{eq:041}  
\end{equation}
and taking into account that (\ref{eq:018}) leads to
\begin{equation}
- \beta z \frac{\partial\nu}{\partial z } =   \int\limits_{-\infty}^{\infty}\left [ \varrho(x) - \theta \varrho \right ] dx,
\label{eq:042}  
\end{equation}
after simple manipulations, we finally obtain
\begin{equation}
\varrho_s =   \int\limits_{-\infty}^{x'} \varrho(x) dx + \int\limits_{x'}^{\infty}\left [ \varrho(x) - \varrho \right ] dx,
\label{eq:043}  
\end{equation}
where $\varrho(x)$ is the number density near the field boundary as a function of the coordinate directed along the field gradient.

As will be seen from the foregoing (\ref{eq:043}) agrees with ordinary equation
\begin{equation}
\varrho_s = - \frac{\partial \gamma}{\partial \mu}.
\label{eq:044}
\end{equation}

Formula (\ref{eq:043}) generalizes the expression for the surface number density \cite{SteckiSokolowski1978,SokolowskiStecki1980,HendersonSwol1984} to the case of arbitrary location of the dividing surface.

Thus, the contribution of the solid phase to the surface number density is defined by $\varrho(x)$, and that of the fluid - by $\varrho(x) - \varrho$.

It is easy to see that actual contribution to (\ref{eq:043}) comes from two regions: the region of the interaction between particles and the surface and the adjacent region of near surface distortions (oscillations) of the number density. (This two regions can intersect.)

For the two cases for a hard solid considered above (section \ref{subsec:03b}), we have
\begin{equation}
\varrho_s =   \int\limits_{0}^{\infty}\left [ \varrho(x) - \varrho \right ] dx
\label{eq:045}  
\end{equation}
and
\begin{equation}
\varrho_s =  -R\varrho + \int\limits_{0}^{\infty}\left [ \varrho(x) - \varrho \right ] dx,
\label{eq:046}  
\end{equation}
where $R$ is the particle radius and the point $x_0$ is placed at the coordinate origin.

Expression (\ref{eq:043}) has a simple meaning. Relying on the obvious relation
\begin{equation}
N =   \int\limits_{V}^{} \varrho(\bm{r}) d\bm{r}
\label{eq:047}  
\end{equation}
and using (\ref{eq:038}) and (\ref{eq:039}), we obtain
\begin{eqnarray}
 N_s =  N - N_b = && \int\limits_{V}^{} \varrho(\bm{r}) d\bm{r} - \varrho(V-V') \label{eq:048}  \\
&&= \int\limits_{V'}^{} \varrho(\bm{r}) d\bm{r} + \int\limits_{V-V'}^{} \left [ \varrho(\bm{r}) - \varrho \right ] d\bm{r}
\nonumber ,
\end{eqnarray}
which, under the assumptions made, is identical to (\ref{eq:043}). Thus, this expression is valid for all fluid densities. Correspondence of (\ref{eq:043}) and (\ref{eq:048}) demonstrates the internal consistency of this approach.

\subsection{\label{subsec:03d}The Henry adsorption constant}

For the Henry constant
\begin{equation}
K_H = \lim_{\varrho\rightarrow 0} \frac{\varrho_s}{\varrho}
\label{eq:049}  
\end{equation}
(\ref{eq:040}) gives
\begin{equation}
K_H =  \int\limits_{-\infty}^{x'} \exp(-\beta v) dx - \int\limits_{x'}^{\infty}[ 1 - \exp(-\beta v)] dx.
\label{eq:050}  
\end{equation}

This expression generalizes the earlier obtained one \cite{SokolowskiStecki1981} to the case of arbitrary localization of the dividing surface and an arbitrary law of the increasing potential when we move inward the body.

The $K_H$ dependence on $x'$ is universal and  can always be extracted explicitly. Indeed, subtracting one value of this constant from another we ascertain that
\begin{equation}
K_H(x') = x' - x''+ K_H(x'').
\label{eq:051}
\end{equation}

Choosing the value $x'' = x_0$ defined by (\ref{eq:033}), and taking into account that
\begin{equation}
K_H(x_0) = 0
\label{eq:052},
\end{equation} 
we obtain 
\begin{equation}
K_H(x') = x' - x_0(T),
\label{eq:053}
\end{equation}
and the problem of $K_H$ calculation reduces to the determination of $x_0$.

From (\ref{eq:033}), it follows that 
\begin{equation}
x_0 = - \int\limits_{-\infty}^{0} \theta(x) dx + \int\limits_{0}^{\infty} \varphi(x) dx. 
\label{eq:054}
\end{equation}

Surface passing through the point $x_0$ by virtue of (\ref{eq:052}) can be called zero adsorption surface.

Now return to basic relations. It follows from (\ref{eq:029}) that  
\begin{equation}
\Omega = -P(V-V') - PAK_H(x') + \nu A,
\label{eq:056}
\end{equation} 
or
\begin{equation}
\gamma =  - PK_H(x') + \nu
\label{eq:057}.
\end{equation}

This equality is one of the most significant expressions in the case under study. It combines the three most important surface parameters - the specific surface $\Omega$-potential $\gamma$, the Henry adsorption constant $K_H$, the nonlinear surface coefficient $\nu$, and one bulk parameter - pressure.

Differentiating (\ref{eq:057}) with respect to the chemical potential and using (\ref{eq:042}), we obtain
\begin{equation}
\varrho_s = \varrho K_H(x') + \int\limits_{-\infty}^{+\infty}\left [ \varrho(x) - \theta \varrho \right ] dx,
\label{eq:058}
\end{equation}
which is another expression for the surface number density for any position of the dividing surface. Here, the parts linear and nonlinear in the bulk density  are clearly separated. This expression is more convenient than its analog, derived on the topological basis for a particular value of  $x'$ and a particular kind of particle-wall potentials \cite{SokolowskiStecki1981}.

Thus, we can consider a partition of the surface quantities on the basis of linearity with the same rights, as on the basis of the localization. 

Indeed, if we define a linear surface number density
\begin{equation}
\varrho_{s,l} = \varrho K_H(x')
\label{eq:059}
\end{equation}
and nonlinear one
\begin{equation}
\varrho_{s,n} = \int\limits_{-\infty}^{+\infty}\left [ \varrho(x) - \theta \varrho \right ] dx,
\label{eq:060}
\end{equation}
then, taking into account (\ref{eq:012}) and (\ref{eq:042}), we get ordinary relations
\begin{equation}
\varrho_{s,l} =  \frac{\partial [ P K_H(x')]}{\partial \mu} 
\label{eq:061}
\end{equation}
and
\begin{equation}
\varrho_{s,n} = - \frac{\partial \nu}{\partial \mu},
\label{eq:062}
\end{equation}
in agreement with (\ref{eq:044}), (\ref{eq:057}).

Naturally
\begin{equation}
\varrho_s = \varrho_{s,l} + \varrho_{s,n},
\label{eq:063}
\end{equation}
and (\ref{eq:059}) - (\ref{eq:063}) define the surface number density in the most general form.

\subsection{\label{subsec:03e}Abruptly changing field}

Let us return to the general equation (\ref{eq:025}) and consider a wall with a hard solid potential. (But we can still consider the case of an arbitrary interparticle potential.) The potential energy of particles in the field of the wall is expressed as
\begin{equation}
v(\bm{r}_i) = v_i = 
	 \left\{ 
			\begin{array}{ll} 
         +\infty & (\bm{r}_i \in V_0)\\   
         0 & (\bm{r}_i \notin V_0).
     	\end{array}  
		\right.
		\label{eq:064}
\end{equation}

In this case, the functions $\varphi_i$ (\ref{eq:023}) take the form of the characteristic functions (\ref{eq:015})
\begin{equation}
\varphi_i \rightarrow \psi^{V_0}_i.
\label{eq:065}
\end{equation}	

Thus, the solid occupies a macroscopic volume $V_0$ inside $V$. As noted above, we assume that the solid is far from the boundaries of the volume $V$.

Equation  (\ref{eq:025})  leads to
\begin{equation}
\Omega^U = -(V-V_0)P(z,T)  +A \nu (z,T) = \Omega_{V-V_0},
\label{eq:066}  
\end{equation}
where for $\nu (z,T)$ before the integration over $x_1$, we have the expression
\begin{eqnarray}
 \nu (z,T) && = k_BT\sum_{t=2}^\infty \frac{z^t}{t!} \label{eq:067} \\
&& \times\int \chi^{V_0}_1 \Big [1 - \prod_{i = 2}^{t} (1 - \psi^{V_0}_i) \Big ] {\cal U}^{(t)}_{1...t} dx_1 d\bm{r}_2 ...d\bm{r}_t
\nonumber,
\end{eqnarray}
where $x_1$ is the coordinate perpendicular to the surface given by the boundary $\psi^{V_0}_1$ and
\begin{equation}
\chi^{V_0}_i = 1 - \psi^{V_0}_i = 
	 \left\{ 
			\begin{array}{ll} 
         0 & (\bm{r}_i \in V_0)\\   
         1 & (\bm{r}_i \notin V_0).
     	\end{array}  
		\right.
		\label{eq:068}
\end{equation}

The function $\psi^v_i$ localizes the $i$-th particle in a volume $v$, and the function $\chi^v_i$ localizes this particle outside this volume.

The first terms of the expression which is analogous to (\ref{eq:067}) were obtained using a diagram technique \cite{Bellemans1962}. The series as a whole was examined in the construction of the open statistical ensemble \cite{zaskulnikov200911a, zaskulnikov201004a}. 

We perform integration over $x_1$ in $\nu$. The functions $\theta$ in (\ref{eq:028}) also take the form of characteristic functions
\begin{equation}
\theta_i \rightarrow  \chi_i = \chi(x_i) = 
	 \left\{ 
			\begin{array}{ll} 
         1 & (x_i < 0)\\   
         0 & (x_i \geq 0),
     	\end{array}  
		\right.
			\label{eq:069}
\end{equation}
and the product $\theta_i$ in the integrand in (\ref{eq:028}) is obviously determined in this case by the maximum $x_i$
\begin{equation}
{f}^{(k-1)}_{2...k} = \int \limits_{- \infty}^{+ \infty} \chi(x_1)  \left [1 - \chi(x_1 + \max \{x_2...x_k\} ) \right ] d x_1.
\label{eq:070}
\end{equation}

Thus, we obtain
\begin{equation}
{f}^{(k-1)}_{2...k} =  \max\{0, x_2 ... x_k\},         
\label{eq:071}
\end{equation}
and for $\nu (z,T)$ we have the expression
\begin{eqnarray}
  \nu (z,T) =&& k_BT\sum_{t=2}^\infty \frac{z^t}{t(t-2)!} \label{eq:072} \\
&& \times \int_{ x_2> \{0, x_3... x_t \} }  x_2  {\cal U}^{(t)}_{0,2...t}  d\bm{r}_2 ...d\bm{r}_t
\nonumber,
\end{eqnarray}
where the symmetry of the Ursell factors under particle permutations is used.

Using the change of variables $\bm{r}'_2 = \bm{r}_2, \bm{r}'_i = \bm{r}_i - \bm{r}_2$, for $i = 3,...t$ and omitting the primes, we arrive at
\begin{equation}
\nu (z,T) = k_BT\sum_{t=2}^\infty \frac{z^t}{t(t-2)!}\int_{ x_2 > 0,\dots x_t > 0}  x_2  {\cal U}^{(t)}_{0,2...t}  d\bm{r}_2 ...d\bm{r}_t
\label{eq:073},
\end{equation}
since the Jacobian of this transformation is unity.

In this case we used the invariance of ${\cal U}^{(t)}$ under particles permutations and under spatial inversion.

For the Henry constant (\ref{eq:050}) in accordance with (\ref{eq:053}) in this case we obtain 
\begin{equation}
K_H = x' - x_{step}, 
\label{eq:074}
\end{equation}
where $x_{step}$ is the step location, which generalizes case (\ref{eq:069}).

In the above, we considered the case where the volume of the solid is limited by condition (\ref{eq:064}), i.e., in fact, we increased the volume of the field by the quantity $AR$. (This corresponds to condition (\ref{eq:033}) for the field of a hard solid.) In some cases, it is reasonable not to include this region in the solid. Then, we obtain the expression
\begin{equation}
\Omega^U = -(V-V_0+AR)P(z,T) + A [R P(z,T) +  \nu (z,T)],
\label{eq:075}  
\end{equation}
in which the bulk part of the solid coincides with the field region, and the surface term has a part linear in number density \cite{hill1959, MeelCharbonneau2009, LairdDavidchak2010}. Naturally, this corresponds to the Henry constant (\ref{eq:074}) $x'- x_{step}=-R$.

In conclusion, a few words about hard spheres. 

A system of hard spheres at a hard wall is an object of a constant attention as a convenient model exercise. For the second surface virial coefficient calculations give minus $\pi D^4/8$, where $D$ is the sphere diameter, and for the third one - minus $149\pi^2 D^7/1680$ \cite{Bellemans1962, SteckiSokolowski1978, MeelCharbonneau2009}.

The calculations, carried out in the present work on the basis of (\ref{eq:067}), (\ref{eq:072}) and independently (\ref{eq:123}) show complete coincidence of results with these data.

\subsection{\label{subsec:03f}Macroscopically smooth field}

We assume here that the characteristic size at which the field undergoes significant changes is macroscopic. Substitution of (\ref{eq:013}) in (\ref{eq:027}) and integration over $\bm{r}_2...\bm {r}_k$ give
\begin{equation}
\nu (z,T) =   \sum_{k=2}^\infty \frac{z^k}{k!} \frac{\partial^{k}P}{\partial z^{k} } \int \limits_{-\infty}^{\infty} \left\{ \theta(x) -  \left[\theta(x)\right]^k \right\} dx  
\label{eq:076}
\end{equation}
or
\begin{equation}
\nu (z,T) =  \int\limits_{-\infty}^{\infty}\left [ \theta P(z)- P(\theta z)  \right ] dx,
\label{eq:077}
\end{equation}
where the terms linear in the activity are added to both integrands.

Substitution of this expression in (\ref{eq:025}) yields the logical result
\begin{eqnarray}
 \Omega^U =&&  -\int\limits_{V}^{} P(\theta z) d\bm{r} = -\int\limits_{V}^{} P(z e^{- \beta u}) d\bm{r}= \nonumber\\
&& -\int\limits_{V}^{} P[\mu - u(\bm{r})] d\bm{r} = -\int\limits_{V}^{} P(\bm{r}) d\bm{r}
\label{eq:078},
\end{eqnarray}
where we used the well-known expression for the chemical potential in the field.

From (\ref{eq:050}), (\ref{eq:057}) we obtain
\begin{equation}
\gamma(x') =  -  \int\limits_{-\infty}^{x'} P(x) dx +  \int\limits_{x'}^{\infty} [P - P(x)] dx.
\label{eq:079}
\end{equation}

Expressions (\ref{eq:077}), (\ref{eq:079}) may be useful in studies of surfaces with macroscopic thickness or in quasi-thermodynamics. The typical structure appearing in (\ref{eq:079}), as will be seen from the following, is universal for $\gamma$.

\subsection{\label{subsec:03g}Exponential field}

Consider the ``wall''-fluid potential of the form
\begin{equation}
v(x) = v_0 \exp(-\frac{x}{l}),
\label{eq:080}
\end{equation}
which used in the numerical calculations \cite{MeelCharbonneau2009}, and deduce the surface functions for it.

Using (\ref{eq:050}) and  the exponential integral expansion 
\begin{equation}
\operatorname{Ei} (\varepsilon) = C + \ln(-\varepsilon) + \sum_{n=1}^{\infty}{\frac{\varepsilon^n}{n! n}}, \; [\varepsilon<0],
\label{eq:081}
\end{equation}
where $C\approx 0.577$ is the Euler constant, in accordance with (\ref{eq:053}), for the Henry constant we obtain
\begin{equation}
K_H = x' - l(\ln \frac{v_0}{k_B T} + C).
\label{eq:082}
\end{equation}

%We see that $x_0$ in (\ref{eq:082}) corresponds to the estimate (\ref{eq:055}).

In a similar way, for the function $f^{(k-1)}_{2...k}$ (\ref{eq:028}) we get
\begin{equation}
f^{(k-1)}_{2...k} = l\ln\left[1+\sum_{n=2}^{k}\exp(-\frac{x_n}{l})\right].
\label{eq:083}
\end{equation}

When the $m$-th particle moves far away from the interface inward to the field region, $x_m<0$ and
\begin{equation}
f^{(k-1)}_{2...k} \approx - x_m,
\label{eq:084}
\end{equation}
which does not disturb the convergence of the integrals in (\ref{eq:027}) due to a stronger decay of the Urcell factors ${\cal U}^{(k)}_{1...k}$.

When $l\rightarrow 0$ expression (\ref{eq:083}) reduces to (\ref{eq:071}), that is easily seen from the behavior of the exponents in (\ref{eq:083}). 

If parameter $l$ takes a macroscopic value, then, using (\ref{eq:013}), we obtain
\begin{equation}
\int {f}^{(k-1)}_{2...k}  {\cal U}^{(k)}_{0,2...k} d\bm{r}_2...d\bm{r}_k =  l \beta \ln k\frac{\partial^{k}P}{\partial z^{k} }  
\label{eq:085}
\end{equation}
and taking into account (\ref{eq:027}), we arrive at
\begin{equation}
\nu (z,T) =   l \sum_{k=2}^\infty \frac{z^k\ln k}{k!}   \frac{\partial^{k}P}{\partial z^{k} },
\label{eq:086}
\end{equation}
naturally, in accordance with (\ref{eq:076}) too.

Despite potential (\ref{eq:080}) is beyond the initial setting of the problem (the hard core presence), the above results are valid at least in some temperature region. To prove this conclusion, we consider a potential that becomes $+\infty$ at some distance far enough from the interface in the region $x<0$ and for the rest is identical to (\ref{eq:080}). This distance is determined by $\theta$ proximity to zero. Such a substitution does not affect the calculation, but will comply with the requirements.

Taking into account the relative simplicity of the obtained relations it is not improbable that the ``wall'' with the present potential may be used along with the hard wall as a convenient model system.

\subsection{\label{subsec:03h}Microinhomogeneous field}

In this section we consider briefly the situation when the field depends on the displacement along the surface. We assume that the simulated solid is crystalline, and thus the field has a periodic surface structure.

Considering the integrals over the coordinates of the first particle that are parallel to the surface, we arrive at their averaging over the surface unit cell.

Thus, $\varphi$ in (\ref{eq:025}) is replaced by
\begin{equation}
\langle \varphi\rangle = \frac{1}{a_{cell}} \int_{cell} \varphi dy dz,
\label{eq:087}
\end{equation}
and ${f}^{(k-1)}_{2...k}$ in (\ref{eq:027}) is replaced by
\begin{eqnarray}
&&\langle f^{(k-1)}_{2...k}\rangle = \int \limits_{- \infty}^{+ \infty} d x_1  \nonumber \\
 && \times \frac{1}{a_{cell}} \int_{cell} \theta(\bm{r}_1)  \Big [1 - \prod_{i = 2}^k\theta(\bm{r}_1 + \bm{r}_i) \Big ]  dy_1 dz_1\label{eq:088} \\
&& = \int \limits_{- \infty}^{+ \infty} \left\langle \theta(\bm{r}_1) \right\rangle  - \langle \theta(\bm{r}_1) \prod_{i = 2}^k\theta(\bm{r}_1 + \bm{r}_i) \rangle d x_1
\nonumber,
\end{eqnarray}
where $a_{cell}$ is the area of the surface cell, and an averaging over $y, z$ at a given depth of $x$ is indicated by triangular brackets.

Equations (\ref{eq:029}) - (\ref{eq:033}) retain their form, with the replacements of $\varphi \rightarrow \langle \varphi \rangle$ and ${f}^{(k-1)}_{2...k}  \rightarrow \langle f^{(k-1)}_{2...k}\rangle$. Expressions (\ref{eq:053}), (\ref{eq:057}) seem to retain their original appearance.

In the case where the contact surface has crystal faces which are differently oriented with respect to the crystal axes, the surface terms must be summed over these faces.

\subsection{\label{subsec:03i}\texorpdfstring{``Local pressure''}{"Local pressure"}}

Integration of (\ref{eq:036}) with respect to the chemical potential  yields
\begin{equation}
\Omega =  - \int\limits_{-\infty}^{\mu} N d\mu',
\label{eq:089}  
\end{equation}
where we leave out an arbitrary function of volume, temperature, and field which arises from the integration, since in the limit of low densities the $\Omega $-potential must tend to zero. Using (\ref{eq:047}), we arrive at the relation
\begin{equation}
\Omega =  - \int\limits_{V}^{} P^{*}(\bm{r}) d\bm{r},
\label{eq:090}  
\end{equation}
where
\begin{equation}
P^{*}(\bm{r}) =  \int\limits_{-\infty}^{\mu} \varrho(\bm{r}) d\mu',
\label{eq:091}  
\end{equation}
which, in this case, is equivalent to
\begin{equation}
\varrho(\bm{r}) = \frac{\partial P^{*}(\bm{r})}{\partial \mu}.
\label{eq:092}
\end{equation}

We see that the number density for the system in the field and $P^{*}$ are linked by exactly the same relation as the usual number density and pressure for a homogeneous system (\ref{eq:012}).

A quantity equivalent to the $P^*$ at a small number density was introduced when  fluctuations in inhomogeneous medium were calculated \cite{StillingerBuff1962} and we shall also call it a ``local pressure'' .

Under the conditions of  the Mayer- type expansion validity, relation (\ref{eq:018}) implies 
\begin{eqnarray}
P^{*}(\bm{r_1},z) &=&k_B T z \theta_1 \label{eq:093} \\
&+& k_B T \theta_1\sum_{n=2}^\infty \frac{z^n}{n!} \int \Big [ \prod_{i = 2}^{n} \theta_i \Big ]{\cal U}^{(n)}_{1...n} d\bm{r}_{2}...d\bm{r}_{n}, \nonumber
\end{eqnarray}
which also resembles the standard expansion of pressure in powers of the activity (\ref{eq:010}). 

It should be noted, however, that $P^{*}(\bm {r})$ and $\varrho(\bm {r})$ do not correspond to one another in a macroscopic sense, i.e., they are not linked by a quasi-thermodynamic relation. It is clear that this would be valid for macroscopically smooth fields, for which one can apply (\ref{eq:013}). In this case, integration over the coordinates yields the pair $P[\mu - u(\bm{r})]$ and $\varrho[\mu - u (\bm{r})]$, related by an ordinary equation of state. Otherwise, the functions $\theta$ distort the integrals of the Ursell factors in (\ref{eq:018}) and (\ref{eq:093}), and we have a different functional relationship.

This consideration, in particular, shows the range of application of the quasi-thermodynamic approach \cite{Tolman1948}.

It is seen from (\ref{eq:091}) that $P^*$ rapidly tends to $P$ when the point of observation moves away from the field region.

We emphasize that equation (\ref{eq:090}) is absolutely general and valid for a system in a force field of arbitrary configuration. In the case of macroscopically smooth fields, it transforms into equation (\ref{eq:078}).

Similarly to the above consideration of the case with the number density,  it is possible to reduce equation (\ref{eq:090}) to
\begin{equation}
\Omega = -P(V - V') - \int\limits_{V'}^{} P^{*} d\bm{r} + \int\limits_{V - V'}^{} (P - P^{*}) d\bm{r}
\label{eq:094}  
\end{equation}
by simple manipulations.

Integration over the surface yields
\begin{equation}
\Omega = -P(V - V') - A \int\limits_{-\infty}^{x'} P^{*}(x) dx + A \int\limits_{x'}^{\infty} [P - P^{*}(x)] dx.
\label{eq:095}  
\end{equation}

Just as it should be, the derivative of (\ref{eq:095}) with respect to the chemical potential gives the expressions for the number of particles: bulk (\ref{eq:039}) and surface (\ref{eq:043}) ones.

For the specific surface $\Omega$-potential $\gamma$ we obtain
\begin{equation}
\gamma(x') =  -  \int\limits_{-\infty}^{x'} P^{*}(x) dx +  \int\limits_{x'}^{\infty} [P - P^{*}(x)] dx.
\label{eq:096}  
\end{equation}

An equation analogous to (\ref{eq:096}) was also obtained when considering fluctuations in piped system \cite{StillingerBuff1962}.

It is obvious that (\ref{eq:096}) ensures that (\ref{eq:034}) is valid.

Differentiating (\ref{eq:096}) with respect to the chemical potential and taking into account (\ref{eq:043}), we can verify the validity of (\ref{eq:044}).

From (\ref{eq:090}) it is clear that the meaning of the bulk density of the $\Omega$-potential can be assigned to $P^*$. However, it will be seen from the following that the ``local pressure'' is not the only candidate for this role.

Using (\ref{eq:026}) and taking into account (\ref{eq:010}) and (\ref{eq:093}), it is easy to obtain the expression for the nonlinear surface coefficient
\begin{equation}
\nu = \int \limits_{-\infty}^{+\infty} [\theta P - P^{*}(x)] dx,
\label{eq:097}
\end{equation}
which defines it in terms of ``local pressure''. Note the similarity of (\ref{eq:097}) and  (\ref{eq:077}).

\subsection{\label{subsec:03j}\texorpdfstring{``Mechanical definition'' of $\gamma$}{"Mechanical definition" of gamma}}

By this definition is meant the procedure of compression and expansion of the system in two different directions with the total conservation of a volume and changes in definite internal surfaces \cite[p.43]{RowlinsonWidom2002}. We apply this procedure in a modified form to the case of a wall/fluid interface.
\begin{figure}[htbp]  
\centering     
    \includegraphics{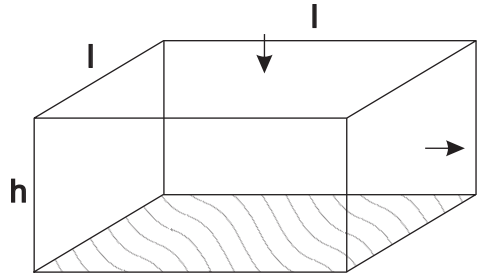} 
   \caption{The system for mechanical definition of the surface $\Omega$-potential of a fluid. The parameters of the lower surface differ from those of the remaining faces.}
   \label{fig:01}
\end{figure}

We assume that the physical surface is inside the system at a microscopically large distance from its boundaries, and, at the same time, this distance is much smaller than the sizes of the system.

The procedure is as follows (see Fig.\ref{fig:01}). First, the upper plane of the system is  shifted downward by $\delta h$ ($\delta h > 0$). The work done on the system at this stage is given by
\begin{equation}
\delta W_1 = \Big [P(l - 2l_t)^2 + 4 l \int \limits_0^{l_t} P_{st_1} d x \Big ] \delta h,
\label{eq:098}
\end{equation}
where $P_{st}$ is the pressure in a narrow strip near the edges of the plane\footnote[2]{For simplicity we assume that all the lengths of the transition regions $l_t, l_ {st}, l^*, l_u$, connected with $P_t, P_{st}, P^*, \theta$ respectively, are identical, unless otherwise stated. Using the averaging integrals, it can be assumed that the integration is performed over the maximum of these lengths.},  which differs from the pressure at the center of the plane (in the bulk). The subscript $s$ indicates that this pressure, generally speaking, does not coincide with the tangential component of the pressure tensor. The fact is that this component is calculated for the effect exerted by the neighboring volumes of a fluid on each other, and in this case we deal with the impact of a fluid on the transverse wall. As this takes place, the whole picture at an atom level is highly distorted.

At the second stage, the system is allowed to expand laterally
\begin{eqnarray}
\delta W_2 =  -  \biggl [ P(l - 2&&l_t)(h - 2l_t) \label{eq:099}   \\
+&& (l+2h) \int \limits_0^{l_t} P_{st_1} d x+  l \int \limits_0^{l_t} P_{st_2} d x \biggr ] \delta l.
\nonumber
\end{eqnarray}

Here, $P_ {st_2}$ is the tangential pressure acting along the bottom edge, and $\delta l > 0$.

The condition of conservation of the volume is not quite trivial: we must assume the constancy of the volume bounded by the planes of the arbitrary localization defined by the coordinate $x'$ measured from the boundaries of the system
\begin{equation}
(l - 2x')^2 \delta h = (l - 2x') (h - 2x') \delta l,
\label{eq:100}
\end{equation}
or, retaining the first-order terms,
\begin{equation}
\delta l = \left ( l + 2x' \frac{l - h}{h} \right ) \frac{\delta h}{h}.
\label{eq:101}
\end{equation}

Then, in the same approximation, we have
\begin{eqnarray}
\delta W &&= \delta W_1 +\delta W_2 \nonumber\\
&&= \delta A_1 \bigg [- \int \limits_0^{x'} P_{st_1} dx + \int \limits_{x'}^{l_t} (P - P_{st_1}) dx \bigg ]  \label{eq:102}\\
&&+ \delta A_2 \bigg [- \int \limits_0^{x'} P_{st_2} dx + \int \limits_{x'}^{l_t} (P - P_{st_2}) dx \bigg ]
\nonumber,
\end{eqnarray}
where
\begin{equation}
\delta A_1 = \left ( -2l + \frac{l^2}{h} \right ) \delta h ~~ \text{and} ~~ \delta A_2 = \frac{l^2}{h} \delta h 
\label{eq:103}
\end{equation}
are changes in the areas of the first and the second types during the process, respectively.

In this example, the shape of a parallelepiped was chosen not by chance: two parameters - $l$ and $h$ - enable $A_1$ and $A_2$ to be varied independently. This allows us to consider only one type of the area to be variable.

By virtue of the assumptions made, the lower and the upper limits of integration can be considered equal to $-\infty$ and $+\infty$. Thus, it can be concluded that, for the fluid, the elementary work is expressed as
\begin{equation}
\delta W =   \delta A \bigg [- \int \limits_{-\infty}^{x'} P_{st} dx + \int \limits_{x'}^{+\infty} (P - P_{st}) dx \bigg ],
\label{eq:104}
\end{equation}
where the tangential pressure $P_ {st}$ acts along the bounding planes.

Since the process is conducted at constant $N$ and $T$, the work should be equal to the change in the free energy $F = F (V, T, N)$. Some complication is the pressure variation during the process, which is, generally speaking, due to adsorption. By varying
\begin{equation}
F = -PV + \gamma A + \mu (N_b + N_S) 
\label{eq:105}
\end{equation}
and taking into account that $\delta N_b = -\delta N_s$, we obtain
\begin{equation}
\delta F = -V \delta P + \gamma \delta A + A \delta \gamma  +  (N_b + N_s) \delta \mu.
\label{eq:106}
\end{equation}

In the right-hand side of the equation all terms are cancelled in pairs, except  the second term, and finally we have
\begin{equation}
\delta F =  \gamma \delta A 
\label{eq:107}
\end{equation}
as it should be if
\begin{equation}
\gamma =   - \int \limits_{-\infty}^{x'} P_{st} dx + \int \limits_{x'}^{+\infty} (P - P_{st}) dx.
\label{eq:108}
\end{equation}

Thus the ``mechanical definition'' of $\gamma$ also gives the same structure, which is already familiar to us (\ref{eq:079}), (\ref{eq:096}).

\subsection{\label{subsec:03k}\texorpdfstring{Tangential force and ``local pressure''}{Tangential force and "local pressure"}}

Let us show a consistency of the approaches (\ref{eq:096}) and (\ref{eq:108}). We differentiate (\ref{eq:105}) with respect to the area using (\ref{eq:096}) for $\gamma$
\begin{equation}
\left ( \frac{\partial F}{\partial A} \right )_{N, V} = -V\frac{\partial P}{\partial A} + \gamma  + A \frac{\partial \gamma}{\partial A} + (N_b+ N_s) \frac{\partial \mu}{\partial A}. \label{eq:109}
\end{equation}

The first term in the right-hand side is cancelled with $N_b \partial\mu/\partial A$ in the last term, and the third term is cancelled with $N_s\partial\mu/\partial A$. This is easily obtained by going over to the derivative with respect to the chemical potential in the integrand and using (\ref{eq:043}), and we finally have
\begin{equation}
\left ( \frac{\partial F}{\partial A} \right )_{N, V} =  -  \int\limits_{-\infty}^{x'} P^{*}(x) dx +  \int\limits_{x'}^{\infty} [P - P^{*}(x)] dx.
\label{eq:110}
\end{equation}

Thus, we can identify (\ref{eq:096}) and (\ref{eq:108}), and immediately obtain
\begin{equation}
\int\limits_{l_t}^{} P_{st} dx =   \int \limits_{l_t}^{} P^* dx
\label{eq:111}
\end{equation}
or
\begin{equation}
 \overline{P_{st}}  =  \overline{P^*},
\label{eq:112}
\end{equation}
where the bar denotes averaging over the transition region. 

In reality, in the absence of clear boundaries of the transition layer, the accuracy of (\ref{eq:112}) is determined by the proximity $P_{st}$ and $P^*$  to zero and $P$ at the boundaries of the averaging interval. Nontrivial part of the equation is determined just by the difference between these quantities and their limits. Thus, the value of the averaging interval in (\ref{eq:112}) is determined by a compromise between these two factors.

\subsection{\label{subsec:03l}Pressure tensor and near-wall distortions}

Let us  now prove a similar equality for the tangential component of the pressure tensor
\begin{equation}
\overline{P_{t}}  =  \overline{P^*}.
\label{eq:113}
\end{equation}

We use the identity of the problem of inhomogeneity in the field with that of the liquid-gas interface for this component \cite{IrvingKirkwood1950}. Let us write the expression
\begin{equation}
P_{t}(x_1) = \varrho_1 k_B T  - \frac{1}{2} \int (y_{2} - y_{1}) \varrho^{(2)}_{12}  \frac{\partial u_{12}}{\partial y_2} d\bm{r}_2,
\label{eq:114}
\end{equation}
which is identical to the expression used in many papers \cite{KirkwoodBuff1949, Harasima1958, OnoKondo1960, SchofieldHenderson1982}. Here $u_{ij}$ is the interaction energy between the $i$-th and the $j$-th particles, $y$ is the coordinate parallel to the surface, and the field gradient, as before, is directed along $x$.

A brief outline of the proof is as follows.

Equations (\ref{eq:093}) and (\ref{eq:114}) are integrated over the coordinates of the first particle, and the integrands are then expanded in a series in the activity using (\ref{eq:020}). Finally, their identity is proved by equating the coefficients at the same powers of $z$. 

A proof of the equality of the integrals of $P_t$ and $P^*$ over volume is given in Appendix \ref{sec:appendc}.

We prove that in the region of homogeneity (no field) $P_t = P^* = P$. Since all the directions in this case are equivalent, for $x$ and $z$ we can write equations similar to (\ref{eq:114}). Composition of these three equations gives
\begin{eqnarray}
&& 3P_{t} = 3 \varrho k_B T  - \frac{1}{2} \int \varrho^{(2)}_{12} \Big [ (x_{2} - x_{1})  \frac{\partial u_{12}}{\partial x_2}  \nonumber\\
&& + (y_{2} - y_{1})  \frac{\partial u_{12}}{\partial y_2} + (z_{2} - z_{1})  \frac{\partial u_{12}}{\partial z_2}  \Big ] d\bm{r}_2.
\label{eq:115}
\end{eqnarray}

Collapsing the expression in brackets, we arrive at the well-known formula for the pressure
\begin{equation}
P_{t} =  \varrho k_B T  - \frac{1}{6} \int \varrho^{(2)}_{12} r_{12}  \frac{\partial u_{12}}{\partial r_{12}}  d\bm{r}_2 = P
\label{eq:116}
\end{equation}
\cite[p.190]{hillstatmeh1987}.

For $P^*$ it is immediately obvious that for $\theta_i = 1, ~i = 1,2, \dots$,  (\ref{eq:093}) reduces to (\ref{eq:010}) or
\begin{equation}
P^* = P
\label{eq:117}
\end{equation}
in the region of homogeneity.

Thus, by virtue of equalities (\ref{eq:116}), (\ref{eq:117}) in the region of homogeneity and the equality of the integrals of $P_t$ and $P^*$ over the volume of the system, we can conclude that (\ref{eq:113}) is valid.

As in the case of $P_{st}$, the transition layer in (\ref{eq:113}) is determined by the proximity $\theta$ to $0$ and $1$ on the edges of the interval.

There are some reasons to believe that the local equality is not valid, i.e., generally speaking,
\begin{equation}
 P_t  \neq    P^*,
\label{eq:118}
\end{equation}
but a rigorous proof of (\ref{eq:118}) for these systems is beyond the scope of this paper. 

Also we shall not deal with the question of pressure tensor ambiguity, because only its zero moment is taken into consideration \cite{SchofieldHenderson1982}.

Thus, we have established the relationship between the tangential pressure and the behavior of near-wall distortions (oscillations)
\begin{equation}
\int \limits_{l_t}^{} \varrho(x) dx = \frac{\partial }{\partial \mu} \int \limits_{l_t}^{}P_t(x) dx
\label{eq:119}
\end{equation}
or
\begin{equation}
\overline{ \varrho} = \frac{\partial \overline{ P_t}}{\partial \mu},
\label{eq:120}
\end{equation}
where the compliance with the macroscopic relation (\ref{eq:012}) also takes place.

Simultaneously we obtained the equality for the tangential force acting on the wall
\begin{equation}
 \overline{P_{st}}  =   \overline{P_t},
\label{eq:121}
\end{equation}
which is not obvious in advance.

Finally, expression (\ref{eq:096}) can be written as
\begin{equation}
\gamma(x') =  -  \int\limits_{-\infty}^{x'} P_{t}(x) dx +  \int\limits_{x'}^{\infty} [P - P_{t}(x)] dx.
\label{eq:122}  
\end{equation}

This expression extends the standard Kirkwood-Buff formula (\ref{eq:001}) to the case of the wall/fluid interface for an arbitrary position of the dividing surface.

\subsection{\label{subsec:03m}\texorpdfstring{Equivalence of ``adsorption'' and ``surface tension'' approaches}{Equivalence of "adsorption" and "surface tension" approaches}}

Equality (\ref{eq:113}) allows us to prove the identity of the ``adsorption'' or ``claster expansion'' approach \cite{Ono1950, hill1959, OnoKondo1960, hillstatmeh1987, Bellemans1962, SokolowskiStecki1980, SokolowskiStecki1981} on the one hand, and the ``surface tension'' or ``pressure tensor'' approach on the other hand \cite{KirkwoodBuff1949, IrvingKirkwood1950, Harasima1958, Navascues1979, HendersonSwol1984, SchofieldHenderson1982}. It suffices to show the equivalence of expressions (\ref{eq:035}) (with $\nu$ in the form  (\ref{eq:026})) and (\ref{eq:122}).

Canceling identical terms, we obtain the equality condition for these expressions in the form
\begin{equation}
\nu = \int \limits_{-\infty}^{+\infty} (\theta P - P_t) dx.
\label{eq:123}
\end{equation}

Comparing (\ref{eq:123}) with (\ref{eq:097}) and taking into account (\ref{eq:113}), we can see that this equality is valid.

In view of (\ref{eq:057}), we have another expression for the specific surface $\Omega$-potential
\begin{equation}
\gamma =  - PK_H(x') + \int \limits_{-\infty}^{+\infty} (\theta P - P_t) dx,
\label{eq:124}
\end{equation}
which, of course, is identical to expression (\ref{eq:122}) but with the terms linear and nonlinear in pressure explicitly separated from each other. Replacing $P_t$ by $P^*$ in (\ref{eq:124}), we return to the approach in terms of ``adsorption'', replacement by $P_{st}$ leads us to the ``mechanical definition'' of $\gamma$.

\subsection{\label{subsec:03n}Near-surface virial expansion}

Equation (\ref{eq:120}) and properties (\ref{eq:113}), (\ref{eq:121}) allow us to construct near-surface analogs of the virial expansion.

As a preliminary let's consider a case of a homogeneous environment. Then we have a system of equations (\ref{eq:010}), (\ref{eq:011}), which can be written as
\begin{equation}
	 \left\{ 
			\begin{array}{cll} 			
        \beta P &=&\displaystyle{z + \sum_{n=2}^{\infty} b_n z^n}  \\   
         \varrho &=& \displaystyle{z + \sum_{n=2}^{\infty} n b_n z^n} 
     	\end{array}  
		\right.
		\label{eq:125},
\end{equation}
where
\begin{equation}
b_n = \frac{1}{n!} \int {\cal U}^{(n)}_{1...n} d\bm{r}_2...d\bm{r}_n
\label{eq:126}.
\end{equation}

We emphasize that the integration in (\ref{eq:126}) is performed over infinite space \cite{zaskulnikov200911a}.

Eliminating $z$, we obtain
\begin{equation}
\beta P = \varrho - \sum_{k=1}^{\infty} \frac{k}{k+1} \beta_k \varrho^{k+1}
\label{eq:127},
\end{equation}
where
\begin{equation}
\beta_k = \sum_{\bm{m}} (-1)^{\sum_j m_j -1} \frac{(k-1+\sum_j m_j)!}{k!} \prod_j \frac{(jb_j)^{m_j}}{m_j!}
\label{eq:128}
\end{equation}
are the so-called irreducible cluster integrals. The summation is performed over all sets of a nonnegative $m_j$, satisfying
\begin{equation}
\sum_{j=2}^{k+1} (j-1) m_j = k
\label{eq:129}
\end{equation}
\cite[p.144]{hillstatmeh1987}.

As is easy to see the condition (\ref{eq:129}) ensures the same dimensionality of different terms in the sum (\ref{eq:128}).

In the presence of the field, we must start from the equations (\ref{eq:093}), (\ref{eq:018}). Averaging them on the transition layer, we obtain the system
\begin{equation}
	 \left\{ 
			\begin{array}{cll} 			
        \beta \overline{P^*}  &=&\displaystyle{ z \overline{\theta} + \sum_{n=2}^{\infty} d_n (z\overline{\theta})^n}  \\   
         \overline{ \varrho} &=& \displaystyle{z \overline{\theta} + \sum_{n=2}^{\infty} n d_n (z\overline{\theta})^n} 
     	\end{array}  
		\right.
		\label{eq:130},
\end{equation}
where
\begin{equation}
d_n = \frac{1}{n!\, l_t \,\overline{\theta}^n} \int \limits_{l_t}  \int \Big [ \prod_{i = 1}^{n} \theta_i \Big ] {\cal U}^{(n)}_{1...n} d x_1 d\bm{r}_2...d\bm{r}_n
\label{eq:131}.
\end{equation}

Here the integration on the coordinate of the $1$-th particle is performed over the transition layer, and on the coordinates of the other particles - over the infinite space.

We can perform integration over $dx_1$ similarly to (\ref{eq:026}). Making the change of variables $x'_1 = x_1,~\bm{r}_i^{'} = \bm{r}_i -\bm {r}_1$ ($i = 2,\dots n$) and using the invariance of ${\cal U}^{(k)}_{1...k}$ under translations, we get
\begin{equation}
d_n = \frac{1}{n!} \int  {h}^{(n-1)}_{2...n} {\cal U}^{(n)}_{0,2...n} d\bm{r}_2...d\bm{r}_n
\label{eq:132},
\end{equation}
where
\begin{equation}
{h}^{(n-1)}_{2...n} = \frac{1}{l_t\,\overline{\theta}^n} \int \limits_{l_t} \theta(x)  \Big [ \prod_{i = 2}^n\theta(x + x_i) \Big ] d x
\label{eq:133}.
\end{equation}

By virtue of the identity of the functional relation $P$ and $\varrho$ in (\ref{eq:125}) via $z$ on the one hand, and $\overline{P}^*$ and $\overline{\varrho}$ in (\ref{eq:130}) via $z\overline{\theta}$ on the other hand, we can immediately write down the analogue of (\ref{eq:127})
\begin{equation}
\beta \overline{P^*} = \overline{ \varrho} - \sum_{k=1}^{\infty} \frac{k}{k+1} \delta_k \overline{ \varrho}^{k+1}
\label{eq:134},
\end{equation}
where
\begin{equation}
\delta_k = \sum_{\bm{m}} (-1)^{\sum_j m_j -1} \frac{(k-1+\sum_j m_j)!}{k!} \prod_j \frac{(jd_j)^{m_j}}{m_j!}
\label{eq:135},
\end{equation}
and condition (\ref{eq:129}) is preserved in its original form.

Finally, taking into account (\ref{eq:113}), (\ref{eq:121}), we can write
\begin{equation}
\beta \overline{P}_t = \overline{\varrho} - \sum_{k=1}^{\infty} \frac{k}{k+1} \delta_k \overline{ \varrho}^{k+1}
\label{eq:136}
\end{equation}
and
\begin{equation}
\beta \overline{P}_{st} = \overline{\varrho} - \sum_{k=1}^{\infty} \frac{k}{k+1} \delta_k \overline{ \varrho}^{k+1}
\label{eq:137}.
\end{equation}

Expression (\ref{eq:136}) and (\ref{eq:137}) is preferable to (\ref{eq:134}): a $P_t$ and $P_{st} $ can be extracted from computer experiments. In addition, $P_t$ can be calculated, because it has a representation in quadratures (\ref{eq:114}).

From the point of view of a nontriviality, averaging interval in (\ref{eq:136}), (\ref{eq:137}) must be chosen minimal, but so that the equalities (\ref{eq:112}), (\ref{eq:113}) are still valid.

Thus, the connection $\delta_k$ with $d_j$ defined by (\ref{eq:135}). For example, the first few relations are 
\begin{eqnarray}
&& \delta_1 = 2 d_2  \nonumber \\
&& \delta_2 = 3 d_3 - 6 d_2^2 \label{eq:138} \\
&& \delta_3 = 4 d_4 -24 d_2 d_3 + \frac{80}{3} d_2^3  \nonumber \\
&& \dots 
\nonumber,
\end{eqnarray}
and they are identical to the corresponding relations between $\beta_k$ and $b_j$ \cite[p. 144]{hillstatmeh1987}.

\section{\label{sec:04}Discussion}

\subsection{\label{subsec:04a}Localization of the dividing surface}

As is seen from the foregoing, many difficulties arising in studies of adsorption are related to the fact that the authors combined two surfaces: system and field. Their separation made in the present contribution by transferring the field into the system immediately clarifies the problem.

It is variant (\ref{eq:031}) which is used in papers \cite[{p.424}]{hillstatmeh1987}, and \cite[section 33]{OnoKondo1960}, as well as in earlier works given therein. This follows from the fact that the derivation procedure employed in these papers assigns the surface part to the liquid volume. However, this was done incorrectly. 

It turns out that  calculation of surface terms calls for integrating beyond the limits of the system. The reason is that surface effects are determined by the difference between inhomogeneous and homogeneous number densities. As for homogeneous number density, it requires that integration be made over correlation volume to both sides symmetric about the surface.  

In the above works integration was rigorously restricted to the system volume, i.e., it did not go beyond the limits of its surface, and therefore could not give a correct result. (Note that this is valid for all terms of the series except the first term).

This approach is also used in current works \cite{RenHsue2003}.  

This question was discussed in the literature, and the above incorrectness was eliminated \cite{SokolowskiStecki1980}. However, the form of the first term remained the same. The dividing surface was fixed at the first atomic layer of a solid: the employed method is unsuitable for other variants \cite{SokolowskiStecki1980}.

Even in the case of correct calculation this variant is inconvenient, since it assigns surface particles to the liquid volume. In particular, given the attraction, i.e., real adsorption, in the first order in density the number of surface particles
\begin{equation}
N_s  \sim   \varrho A\int\limits_{l_t}^{} [ \exp(-\beta v) - 1] dx
\label{eq:139}
\end{equation}
may become negative for definite forms of the interaction potential, which is seen from (\ref{eq:139}) and (\ref{eq:043}). 

It can easily be understood, indeed, actually in expressions (\ref{eq:031}), (\ref{eq:139}) we assign the entire transition region to a homogeneous liquid, thus correction should be subtracted from the number of surface particles. 

Variant (\ref{eq:030}) is preferable in the presence of real adsorption, since it agrees with the common notion of particles adsorbed on a solid surface. In the first order in $\varrho$ it always gives a positive value of the number of surface particles. As follows from (\ref{eq:040}), in this case 
\begin{equation}
N_s \sim   \varrho A\int\limits_{l_t}^{} \exp(-\beta v) dx, 
\label{eq:140}
\end{equation}
i.e., always positive. 

Here particles are divided into two parts: particles belonging to the homogeneous liquid and those belonging to the transition layer; they are not taken into account twice as in (\ref{eq:139}). In other words, we add them in (\ref{eq:140}) as external relative to the liquid, and in (\ref{eq:139}) they are subtracted as internal.

The attempt to directly identify these variants leads to two definitions of adsorption, ``Gibbs'' and ``mechanic'', as well as to the introduction of the ``effective potential'' \cite{hill1959}. 

Variant (\ref{eq:032}) is inapplicable in the presence of real adsorption, since it assigns molecules adsorbed on the surface to the liquid volume, i.e., it is obviously a workaround, artifice in this case. Indeed, the presence of pronounced negative potential near the surface, as is readily seen from (\ref{eq:054}), shifts the surface position $x_0$ deep into the solid.  

This variant is good when the potential of the surface is close to that of a hard solid or, speaking more generally, in the absence of a pronounced adsorption.

\subsection{\label{subsec:04b}Surface number density}

At arbitrary number density when the dividing surface is placed on external boundary of the field, expression (\ref{eq:043}) takes the form 
\begin{equation}
\varrho_s =   \int\limits_{l_u}^{} \varrho(x) dx + \int\limits_{0}^{\infty}\left [ \varrho(x) - \varrho \right ] dx,
\label{eq:141}
\end{equation}
where the external boundary of the field in the second integral is placed at the origin of coordinates.  

Here the two terms are defined most sharply. The first term in the right-hand side of (\ref{eq:141}) corresponds to the adsorption in a classical understanding as particles residing on the surface of a solid, is linear in the number density, and is defined by the region of direct action of the wall field on fluid particles.  

The second term is quadratic in the number density (which follows from (\ref{eq:018})), and is specified by indirect action of the field by its transfer through the layer of adsorbed particles. It is related to interparticle interaction and geometric displacement of particles from the region of ``dead'' volume near the surface. 

As is should be,  for the hard solid only the second term works in (\ref{eq:141}), and it reduces to (\ref{eq:045}), while for macroscopically smooth field - only the first term, and (\ref{eq:141}) transforms into

\begin{equation}
\varrho_s =   \int\limits_{l_t}^{} \varrho(x) dx,
\label{eq:142}
\end{equation}
which, as is easily seen, agrees with (\ref{eq:078}). 

The second term in (\ref{eq:141}) may be called ``indirect adsorption'' unlike ordinary one (the first term) that may be called ``direct adsorption''. Though, these two terms are not strictly separated. 

Some remarks concerning the dividing surface. Differentiating expression (\ref{eq:066}) with respect to the chemical potential, one can see that in this case the most natural choice is the Henry constant of adsorption is equal to zero. Exactly this choice of the dividing surface is used in \cite{Bellemans1962}. Recall that it corresponds to the inclusion of near-surface volume $AR$, inaccessible to fluid particles, in the solid volume.

On the contrary, in a great number of computational and other works it is common practice to assign ``dead'' volume to a fluid one, which results, as is evident from (\ref{eq:075}), in additional negative term $-R$ in the above-mentioned constant \cite{LairdDavidchak2010}. Note that results of paper \cite{LairdDavidchak2010} for surface number density (both theoretical and those obtained by computer simulation) agree with (\ref{eq:075}). This is seen from the comparison between the derivative of this expression with respect to chemical potential and the expansion of two variants of Percus–Yevick equation into a series in density \cite{LairdDavidchak2010}. 

In conclusion note that unlike the case of the liquid-gas equilibrium system, in the variant at hand imposing the condition $N_s = 0$, as is seen from (\ref{eq:043}), results in that the position of the dividing surface becomes dependent on the number density. Probably, in the given case there is no physical meaning of such a separation. Instead, in some situations one can use condition (\ref{eq:033}).

\subsection{\label{subsec:04c}Tangential force vs pressure tensor}

As was mentioned above, we cannot identify the specific tangential force and tangential component of the pressure tensor just from general considerations. In other words, we cannot a priori believe that $P_{st} = P_t$ and even that $\overline{P}_{st} = \overline{P}_t$ . There are two reasons for it. 

First, $P_t$ is introduced for the conditions, which ensure a uniformity along the surface. It specifies the interaction of the neighboring volumes of the fluid. On the contrary, entering of $P_{st}$ implies the presence of a transverse wall, i.e., the involvement of a corner structure. It is clear, that in this case particles distribution has nothing in common with that in the previous case.

Second, $P_t$ is not the force, but the time derivative of momentum \cite{IrvingKirkwood1950}. It consist of the force and the kinematic terms. The special theorem (or theorems) is needed to certify that this momentum flow is equal to the corresponding part of the force.

The proof that $\overline{P}_{st} = \overline{P}_t$ given in the present paper is one of such theorems.

\subsection{\label{subsec:04d}Adsorption and surface tension}

Obviously expression (\ref{eq:044}) is of a universal nature. So the surface number density (which just specifies an adsorption in a broad sense) is closely connected with the specific surface $\Omega$-potential $\gamma$.

As follows from the ``mechanical definition'' of $\gamma$ (Section \ref{subsec:03j}) and theorem (\ref{eq:121}), the surface forces are determined by the same value.

Thus, we may state that the surface tension (also in a broad sense) for the case in point is closely connected with adsorption.

Therefore, the proof of the identity of the two approaches denoted as ``adsorption'' and ``surface tension'' performed in the present contribution is primarily related not to these notions but to methodically different directions of the research. The first approach initially appeals to the systems in the equilibrium, while the second one starts from the analysis of the force dynamics \cite{KirkwoodBuff1949}.

On the other hand, the reasonability of using the terms ``adsorption'' and ``surface tension'' in the given context is undoubted, due to the features of the evolution of the above-mentioned approaches.

Note that ``direct'' and ``indirect'' adsorptions are not strictly distinguished, since both terms are involved in the tangential pressure. Besides, as is seen from the comparison between (\ref{eq:043}) and (\ref{eq:058}), both of them, generally speaking, give a contribution to the nonlinear surface coefficient $\nu$.

\subsection{\label{subsec:04e}Surface thermodynamic potentials}

Equation (\ref{eq:034}) coincides with the expression for $\Omega$-potential of the liquid/gas system \cite{OnoKondo1960}. It is this potential which is most natural for the phase boundary. However, for the liquid/gas interface we can always choose the position of the dividing surface such that the surface number density goes to zero. In this case the surface  $\Omega$ - potential coincides with the corresponding free energy \cite{OnoKondo1960}. 

As already mentioned, in the case under consideration the surface number density does not go to zero, so $\gamma(x')$ is no longer a specific surface free energy. Indeed, 
\begin{equation}
F = \Omega + \mu N = -P(V - V')+ \gamma(x') A + \mu (N_b + N_s) = F_b + F_s,
\label{eq:143}
\end{equation}
where
\begin{equation}
F_b =  -P(V - V') + \mu N_b
\label{eq:144}  
\end{equation}
and
\begin{equation}
F_s = A [\gamma(x') + \mu \varrho _s]
\label{eq:145}  
\end{equation}
are volume and surface free energies, respectively.

\section{\label{sec:05}Summary}

\begin{enumerate}

	 \item The general expression of the $\Omega$-potential (grand potential) of the system with locally introduced force field is derived (\ref{eq:025}). This technique makes it possible to vary the position of the dividing surface (\ref{eq:029}). 	 
	 \item Volume and surface terms of the $\Omega$-potential cannot be unambiguously separated, and their interpretation is specified by the form of the solid-fluid interaction potential (\ref{eq:030}) - (\ref{eq:032}). 		
		\item In particular, depending on the situation, the surface term can begin both with linear and with quadratic term in the activity (\ref{eq:030}), (\ref{eq:032}), which corresponds to the presence or to the absence of adsorption in ordinary understanding. 		
		 \item Solid phase contribution to the surface number density is defined by $\varrho(x)$, and fluid contribution - by the difference $\varrho(x) - \varrho$ (\ref{eq:043}). 		 
		 \item The general expression for the specific surface $\Omega$-potential $\gamma$, composed of the terms linear and nonlinear in pressure is obtained (\ref{eq:057}). The linear term is determined by the product of pressure and the Henry adsorption constant (\ref{eq:050}), and depends on the location of the dividing surface. The nonlinear term (the nonlinear surface coefficient - $\nu$) has several universal representations (\ref{eq:026}), (\ref{eq:027}), (\ref{eq:097}), (\ref{eq:123}).			 		 
		 \item The surface number density, similarly to the $\gamma$, breaks up into the term linear in the bulk density depending on the dividing surface position, and universal nonlinear term (\ref{eq:058}).		 
		 \item The Henry constant of adsorption depends on the position of the dividing surface, and is determined by the universal expression $x' - x_0$ (\ref{eq:053}). 				
		 \item The expression for the nonlinear surface coefficient $\nu$ is obtained, in the form of a series in powers of the activity - the analog of the Mayer series for pressure (\ref{eq:027}). The coefficients of the series are given as integrals of the products of Ursell factors and simple multipliers, that depend on external potential (\ref{eq:028}). 		 
		 \item For the hard solid case (section \ref{subsec:03e}) the coefficients of the series are the first moments of Ursell factors taken over certain regions (\ref{eq:073}). 		 
		 \item For macroscopically smooth field of arbitrary configuration (section \ref{subsec:03f}) the expression obtained for $\nu$ (\ref{eq:077}) agrees with thermodynamically reasonable expression for $\Omega$-potential (\ref{eq:078}).			 
		 \item Probably, the system with exponential wall-fluid potential, can be used as a convenient model for the calculation of surface terms (section \ref{subsec:03g}).			 
		  \item The expressions for $\nu$ and $\Omega$-potential of microscopically inhomogeneous field of periodic surface structure are deduced (section \ref{subsec:03h}).	  		  
		  \item The ``surface tension'' approach produces the expression for $\nu$ in quadratures (\ref{eq:123}), which permits to operate with $\gamma$ as a whole (\ref{eq:124}).				    
		  \item The ``adsorption'' approach produces the expression for $\nu$ in quadratures, when ``local pressure'' (\ref{eq:097}) is used.		  		  
		  \item On  average over the transition region near the surface the tangential component of pressure tensor  (\ref{eq:114}) coincides with the pressure exerted on the transverse wall (\ref{eq:098}) and with the ``local pressure'' (\ref{eq:093}). This fact together with arbitrary position of the separating surface allows one to consider new versions of the Kirkwood-Buff formula (\ref{eq:096}), (\ref{eq:108}), (\ref{eq:122}). 			  
		  \item The $\Omega$-potential of a statistical system in a force field of arbitrary configuration is defined by the integral of ``local pressure'' over the entire volume (\ref{eq:090}), (\ref{eq:093}). In view of the previous item, it is also defined by analogous integrals of $P_t$ and $P_{st}$, when these quantities are valid.		 		  
		  \item On average over the transition region the relation between the near-surface number density and tangential component of pressure tensor corresponds to ordinary macroscopic relation between number density and pressure (\ref{eq:120}).		 		  
		  \item For the problems under study, ``adsorption'' and ``surface tension'' that evolved as independent lines of investigations are completely equivalent in their common domain of existence (section \ref{subsec:03m}). 
		  \item The near-surface virial expansion (section \ref{subsec:03n}) determines the equation of state of near boundary   ``two-dimensional'' fluid. 
		 		 		 
\end{enumerate}

%\appendix

\begin{appendices}

\numberwithin{equation}{section}

\section{\label{sec:appenda}Factors}

\subsection{\label{subsec:appenda1}Ursell factors}

Ursell factors may be defined by the equality:  
\begin{eqnarray}
{\cal U}^{(k)}_{1...k} &=& \sum_{\{\bm{n}\}}(-1)^{l-1}(l-1)!\prod_{\alpha = 1}^l \exp[-\beta U^{k_\alpha}(\bm{n}_\alpha)],  \nonumber \\
1 & \leq & ~ k_\alpha \leq k, ~ \sum_{\alpha = 1}^l k_\alpha = k, ~ \exp(-\beta U^{1}) = 1,
\label{eq:a01}
\end{eqnarray}
where by  $\{\bm{n}\}$ we denote some partition of the given set of $k$ particles with the coordinates $\bm{r}_1,...\bm{r}_k$ into disjoint groups $\bm{n}_\alpha$, $l$ is the quantity of groups of a particular partition, $k_\alpha$ is the size of the group with the number $\alpha$. The sum is taken over all possible partitions, and the meaning of the condition $\exp(-\beta U^{1}) = 1$ is evident: unit groups in  the given case make no contribution to the products. 

For example, the first several  ${\cal U}^{(k)}_{1...k}$ are 
\begin{eqnarray}
{\cal U}^{(1)}_{1}\mkern 9mu &=& 1 \label{eq:a02} \\
{\cal U}^{(2)}_{1,2} \mkern 8mu &=& \exp(-\beta U^{2}_{1,2}) - 1 \nonumber \\
{\cal U}^{(3)}_{1,2,3} &=& \exp(-\beta U^{3}_{1,2,3}) - \exp(-\beta U^{2}_{1,2}) \nonumber \\
 &-& \exp(-\beta U^{2}_{1,3}) - \exp(-\beta U^{2}_{2,3}) + 2 \nonumber \\
\dots \nonumber
\end{eqnarray}

As is well-known, one of the generating functions for Ursell factors is the logarithm \cite{percus1964}. However, the division of the series in (\ref{eq:009}) for the case $k=1$ by $\Xi_V$ in  form (\ref{eq:007}) shows that there exists the second generating function - fractional - in the form of the series ratio (\ref{eq:009}). This follows from the familiar expansion of the number density for homogeneous system (\ref{eq:011}) that is valid far from the system boundaries. 

In Appendix \ref{subsec:appendb1} the existence of fractional generating function will be proved.

\subsection{\label{subsec:appenda2}Partial localization factors}

As far as we know, these functions were introduced for the first time in \cite{RuelleStatmeh1969}.

Some particles involved in these functions do not cause the decay when moving away (the delocalized group), and some of them - do (the localized one). 

Introduce the notation
\begin{equation}
{\cal B}^{(m,k)}_{1...m+k},
\label{eq:a03}
\end{equation}
where the superscripts $m$ and $k$ define the quantity of delocalized and localized particles, respectively ($m = 1,2,3,\dots, k = 0,1,2,\dots$). The subscripts denote the coordinates of particles, with the first $m$ particles being considered delocalized, and the rest - localized.

These functions are similar in structure to the Ursell factors of  $k+1$-th rank, with the proviso that in the construction by the type of (\ref{eq:a01}) the first $m$ particles (delocalized) are treated as a single compound particle. In other words, define ${\cal B}^{(m,k)}_{1...m+k}$ by the equality 
\begin{eqnarray}
{\cal B}^{(m,k)}_{1...m+k} &=& \sum_{\{\bm{n}\}}(-1)^{l-1}(l-1)!\label{eq:a04}\\
&\times& \prod_{\alpha = 1}^l \exp[-\beta U^{k_\alpha+(m-1)\delta_{\alpha\tau}}(\bm{n}_\alpha)],  \nonumber \\
1  \leq&  k_\alpha& \leq k+1; ~~ \sum_{\alpha = 1}^l k_\alpha = k+1; ~~ \exp(-\beta U^{1}) = 1,\nonumber 
\end{eqnarray} 
where designations are analogous to (\ref{eq:a01}) on condition that summation is taken over all possible partitions of the set of $k+1$ particles among which one particle is compound. $\delta_{\alpha\tau}$ is the Kronecker delta, and $\tau$ is the number of the group involving a compound particle. 

The generating function for ${\cal B}^{(m,k)}_{1...m+k}$  is the distribution function $\varrho^{(m)}_{G,1...m}$ of GCE type. Expanding the partition function $\Xi_V$ in (\ref{eq:009}) and dividing the series, we obtain (\ref{eq:014}). The proof of this relation is given in Appendix \ref{subsec:appendb1}. 

${\cal B}^{(m,k)}_{1...m+k}$ that are the first in localized group: 
\begin{eqnarray}
{\cal B}^{(m,0)}_{1...m} \mkern 12mu &=& \exp(-\beta U^{m}_{1...m})\label{eq:a05} \\
{\cal B}^{(m,1)}_{1...m+1} &=& \exp(-\beta U^{m+1}_{1...m+1}) - \exp(-\beta U^{m}_{1...m})\nonumber \\
{\cal B}^{(m,2)}_{1...m+2} & = & \exp(-\beta U^{m+2}_{1...m+2}) -  \exp(-\beta U^{m+1}_{1...m+1})\nonumber \\
 &-& \exp(-\beta U^{m+1}_{1...m,m+2}) - \exp(-\beta U^{m}_{1...m}) \nonumber \\
 &\times& \exp(-\beta U^{2}_{m+1,m+2})  + 2\exp(-\beta U^{m}_{1...m}) \nonumber \\
\dots \nonumber
\end{eqnarray}
and in delocalized one:
\begin{equation}
{\cal B}^{(1,k-1)}_{1...k} = {\cal U}^{(k)}_{1...k}
\label{eq:a06},
\end{equation}
including, for homogeneous medium 
\begin{equation}
{\cal B}^{(1,0)}_{1} = 1
\label{eq:a07}.
\end{equation}

It is obvious from (\ref{eq:a05}), and (\ref{eq:a06}) that partial localization factors generalize the concepts of Boltzmann factors and Ursell factors, including them as the limiting cases.  

For factors ${\cal B}^{(m,k)}_{1...m+k}$ a number of recurrence relations hold, which ensure the existence of various physical links; some of them are given in Appendix \ref{sec:appendb}.

\section{\label{sec:appendb}\texorpdfstring{Recurrence relations for ${\cal B}^{(m,k)}_{1...m+k}$}{Recurrence relations for B(m,k)1...m+k}}

Many relations and operations of statistical mechanics are provided by definite classes of recurrence relations for ${\cal B}^{(m,k)}_{1...m+k}$. For brevity, we shall say that an operation generates a recurrence relation or a class. In this contribution only some of them will be considered. More detailed information can be found in \cite{zaskulnikov201004a}.

\subsection{\label{subsec:appendb1}Correspondence to the definition}

Equating expressions (\ref{eq:009}) and (\ref{eq:014}), expanding the series for $\Xi_V$, and performing multiplication of the series, we arrive at 
\begin{equation}
{\cal B}^{(m,k)}_{1...m+k} = {\cal B}^{(m+k,0)}_{1...m+k} - \sum_{n=1}^k \sum_{\text{samp}} {\cal B}^{(n,0)}_{1...n}{\cal B}^{(m,k-n)}_{n+1...m+k}
\label{eq:b01},
\end{equation}
where $m \geq 1$, $k \geq 1$, and the internal sum is taken over samplings of localized particles only ($n$ from $k$). Relation (\ref{eq:b01}) is proved either by direct exhaustion of partitions in accordance with (\ref{eq:a04}), or by repeatedly substituting the expression for ${\cal B}^{(m,k)}_{1...m+k}$ in the right-hand side of (\ref{eq:b01}). 

Paper \cite{percus1964}  gives another recurrence relation for Ursell functions which in terms of ${\cal B}^{(m,k)}_{1...m+k}$ looks like
\begin{equation}
{\cal B}^{(1,k)}_{1...k+1} = {\cal B}^{(k+1,0)}_{1...k+1}- \sum_{n=1}^{k} \binom{k}{n} \left [ {\cal B}^{(n,0)}_{1...n} {\cal B}^{(1,k-n)}_{n+1...k+1}\right ]_{\text{perm}}
\label{eq:b02}
\end{equation}
where $k \geq 1$, and square brackets denote averaging over permutations of particles. 

In principle, this is the same equation (\ref{eq:b01}) at $m=1$ but written in a symmetric form. Note that in the given case there is no necessity in symmetrization - equation  (\ref{eq:b01}) holds rigorously in asymmetric form as well. Nevertheless, (\ref{eq:b01}) can also be brought into a symmetric form, however, this calls for averaging over permutations of both the sums and the left-hand side of the equation.

\subsection{\label{subsec:appendb2}BBGKI equation}

In the presence of external field this set of linking equations may be written as \cite[p.205]{hillstatmeh1987}
\begin{eqnarray}
{\nabla}_1\varrho^{(m)}_{1...m} = - \beta && \varrho^{(m)}_{1...m} \Big ({\nabla}_1 v_1 + \sum_{i=2}^{m }{\nabla}_1 u_{1i} \Big ) \nonumber \\ && - \beta  \int  \varrho^{(m+1)}_{1...m+1} {\nabla}_1 u_{1m+1} d\bm{r}_{m+1}. \label{eq:b03}
\end{eqnarray}

This relation, after substituting expressions (\ref{eq:020}) in it and equating coefficients at equal powers of $z$, generates the differential recurrence relation
\begin{eqnarray}
{\nabla}_1{\cal B}^{(m,n)}_{1...m+n} = -&& \beta  {\cal B}^{(m,n)}_{1...m+n} \sum_{i=2}^{m }{\nabla}_1 u_{1i} \nonumber \\ -&& \beta \sum_{i=m+1}^{m+n }{\cal B}^{(m+1,n-1)}_{1i..m+n}{\nabla}_1 u_{1i},
\label{eq:b04}
\end{eqnarray}
which we use to prove (\ref{eq:113}), (\ref{eq:119}). In the second term on the right all permutations of particles of the localized group ($n$) with the second particle are exhausted.

\section{\label{sec:appendc}\texorpdfstring{Proof of $\langle P_t\rangle = \langle P^*\rangle$}{Proof of <Pt> = <P*>}}

\begin{figure}[htbp]  
  \includegraphics{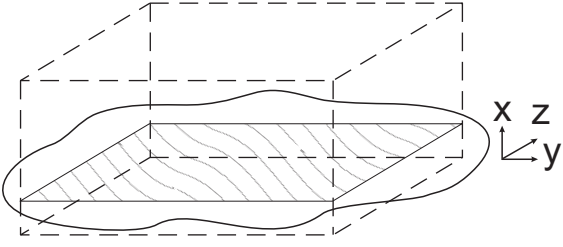}
   \caption{The system used to prove $\langle P_t\rangle = \langle P^*\rangle$. Fluid is at the top of the interface.}
   \label{fig:02}
\end{figure}

To prove this equality, consider the system inside the volume shaped as rectangular parallelepiped strongly stretched along the $y$ axis (see Fig.\ref{fig:02}). The system includes the interface plane, and one of the system boundaries lies fairly deep inside the field region. Wall/fluid interaction is still defined by the functions $\theta$.

Let us integrate expressions (\ref{eq:093}) and (\ref{eq:114}) with respect to the coordinates of the first particle. The integration is performed over the whole system volume denoted by dashed line in Fig. \ref{fig:02}. We put that this volume is defined by the characteristic function $\psi_1$.  

Expand the integrands into a series in the activity using (\ref{eq:020}). Equating the coefficients at equal powers of $z$, we have 
\begin{eqnarray}
&&  k_B T \int \psi_1 \bigg [ \prod_{i = 1}^{k+2} \theta_i \bigg ] {\cal B}^{(1,k+1)}_{1...k+2} d\bm{r}_1 ...d\bm{r}_{k+2} \label{eq:c01}\\
&& = \frac{(k+2)}{2} \int \psi_1 \bigg [ \prod_{i = 1}^{k+2} \theta_i \bigg ] (y_2 - y_1) {\cal B}^{(2,k)}_{1...k+2} \frac{\partial u_{12}}{\partial y_2} d\bm{r}_1 ...d\bm{r}_{k+2},
\nonumber 
\end{eqnarray}
where $k = 0,1,\dots$. Here we equated the coefficients at $z^{k+2}$. So, it is necessary to prove this equality. 

The kind of the recurrence relation (\ref{eq:b04}) required to prove (\ref{eq:c01}) is of the form 
\begin{equation}
- k_B T \frac{\partial}{\partial y_2}{\cal B}^{(1,k+1)}_{1...k+2} = \sum_{i=1,i\neq2}^{k+2 }{\cal B}^{(2,k)}_{2i..k+2}\frac{\partial u_{2i}}{\partial y_2} .
\label{eq:c02}
\end{equation}

Multiplying (\ref{eq:c02}) by $(y_2 - y_1)\psi_1\theta_1...\theta_{k+2}$ and integrating over all coordinates, we obtain using the symmetry under permutations inside localized and delocalized groups of ${\cal B}^{(m,n)}$ 
\begin{eqnarray}
 - k_B T  \int \psi_1 \bigg [ \prod_{i = 1}^{k+2} \theta_i \bigg ]&& (y_2 - y_1) \frac{\partial}{\partial y_2}{\cal B}^{(1,k+1)}_{1...k+2} d\bm{r}_1 ...d\bm{r}_{k+2} \nonumber \\
&& =  I_1 + k I_2 +k I_3, \label{eq:c03} 
\end{eqnarray}
where
\begin{equation}
I_1=\int  \psi_1  \bigg [ \prod_{i = 1}^{k+2} \theta_i \bigg ](y_2 - y_1) {\cal B}^{(2,k)}_{2,1...k+2} \frac{\partial u_{21}}{\partial y_2} d\bm{r}_1 ...d\bm{r}_{k+2},
\label{eq:c04}
\end{equation}
\begin{equation}
I_2=\int  \psi_1 \bigg [ \prod_{i = 1}^{k+2} \theta_i \bigg ](y_2 - y_3) {\cal B}^{(2,k)}_{2,1...k+2} \frac{\partial u_{21}}{\partial y_2} d\bm{r}_1 ...d\bm{r}_{k+2},
\label{eq:c05}
\end{equation}
\begin{equation}
I_3= \int (\psi_3 - \psi_1)\bigg [ \prod_{i = 1}^{k+2} \theta_i \bigg ] (y_2 - y_3) {\cal B}^{(2,k)}_{2,1...k+2} \frac{\partial u_{21}}{\partial y_2} d\bm{r}_1 ...d\bm{r}_{k+2}.
\label{eq:c06}
\end{equation}

Here in the integrals $I_2$ and $I_3$ the change of variables $\bm{r}_1 \leftrightarrow \bm{r}_3$ is performed. 

The integral in the left-hand side of equality (\ref{eq:c03}) may be taken by parts, since here factors $\theta_i$ depend solely on $x_i$. Wherein, due to locality of Ursell factors, and factors ${\cal B}^{(1,n)}$ are exactly the factors of this kind, the integral term goes to zero. Thus, 
\begin{eqnarray}
 - k_B T && \int  \psi_1 \bigg [ \prod_{i = 1}^{k+2} \theta_i \bigg ] (y_2 - y_1) \frac{\partial}{\partial y_2}{\cal B}^{(1,k+1)}_{1...k+2} d\bm{r}_1 ...d\bm{r}_{k+2} \nonumber \\
&& = k_B T \int \psi_1 \bigg [ \prod_{i = 1}^{k+2} \theta_i \bigg ] {\cal B}^{(1,k+1)}_{1...k+2} d\bm{r}_1 ...d\bm{r}_{k+2} \label{eq:c07}. 
\end{eqnarray}

It is readily seen that 
\begin{equation}
I_1 = 2 I_2.
\label{eq:c08}
\end{equation}

For this purpose, calculate the difference
\begin{equation}
I_1 - I_2 = \int  \psi_1  \bigg [ \prod_{i = 1}^{k+2} \theta_i \bigg ] (y_3 - y_1) {\cal B}^{(2,k)}_{2,1...k+2} \frac{\partial u_{21}}{\partial y_2} d\bm{r}_1 ...d\bm{r}_{k+2}.
\label{eq:c09}
\end{equation}

Making the change of variables $\bm{r}_1 \leftrightarrow \bm{r}_2$, we see that the right-hand side of (\ref{eq:c09}) coincides with $I_2$, so we have (\ref{eq:c08}). 

Finally, consider the integral $I_3$. Obviously, the factor $(\psi_3 - \psi_1)$ is nonzero only in the case where the first and the third particles are on different sides of the system boundary. It is seen better from the expression
\begin{equation}
\psi_3 - \psi_1 = \psi_3\chi_1 - \psi_1\chi_3,
\label{eq:c10}
\end{equation}
where we used the identity $\psi_i + \chi_i \equiv 1$.

When the $1$-st or the $3$-rd particle moves away from the boundary of the system, the integrand decays rapidly owing to local character of the product ${\cal B}^{(2,k)}_{2,1...k+2}\partial u_{21}/\partial y_2$. Hence, the integral is proportional to the surface of the parallelepiped at hand, but not all of its faces make a contribution to $I_3$.

Consider the top face of the system, where due to remoteness from the interface we can consider that $\theta_i = 1, i = 1,2,3\dots$. Making the substitution (\ref{eq:c10}), we arrive at the conclusion that condition $I_3 = 0$ demands in this case that
\begin{eqnarray}
\int \psi_3\chi_1 &&(y_2 - y_3) {\cal B}^{(2,k)}_{2,1...k+2} \frac{\partial u_{21}}{\partial y_2} d\bm{r}_1 ...d\bm{r}_{k+2} \label{eq:c11} \\
&&= \int \psi_1\chi_3 (y_2 - y_3) {\cal B}^{(2,k)}_{2,1...k+2} \frac{\partial u_{21}}{\partial y_2} d\bm{r}_1 ...d\bm{r}_{k+2}.
\nonumber
\end{eqnarray}

However, obviously, equation (\ref{eq:c11}) is true by symmetry when the $y$ axis is parallel to the plane defined by the function $\psi$. (To prove this it suffices to consider the mirror reflection operation with its mirror-plane coinciding with the upper face of the system.) Thus, we may conclude that contribution to $I_3$ from the whole region of homogeneity above the interface equal zero except the side faces $y = const$. In the region of homogeneity below the interface this contribution goes to zero owing to $\theta_i\rightarrow 0$.

So, only the side faces $y = const$ and two narrow strips on the faces $z = const$ make contributions to $I_3$. The contribution of these strips is of higher order of smallness and may be neglected. Thus, we have the estimation
\begin{equation}
\frac{I_3}{S_{zy}} =  o(S_{zy}).
\label{eq:c12}
\end{equation}

The contribution of $I_3$ tends to zero with the system length (along the $y$ axis) tending to infinity. Here $S_{zy}$ is the area of the face $zy$. 
 
In view of (\ref{eq:c08}) and (\ref{eq:c12}), equation (\ref{eq:c03}) gives (\ref{eq:c01}).

Note, that the selection of pressure tensor representation in the form of (\ref{eq:114}) is not accidental. We could omit the term $y_1$ in the integrand, but it will lead to poor localization of functions and complicate the proof.

In a similar way, but easier, it may be proved that relation (\ref{eq:116}) corresponds to expansions (\ref{eq:010}) and (\ref{eq:011}), (\ref{eq:014}).

\end{appendices}

\begin{strip}
\center{---------------------------------------------------------------------------------------------}
\end{strip}

\tiny 
\raggedleft
VZ, 07.12.2011, v54

\end{document}